\documentclass[11pt]{article}
\usepackage{amsmath,amssymb,amsfonts} 
\usepackage{latexsym}   
\usepackage{hyperref}
\usepackage{epsfig}
\usepackage{pgf,tikz}
\def\qed{\hfill$\Box$}
\begin{document}
\title{\vspace{-6ex}\bf Hidden supersymmetry and quadratic deformations of the 
space-time conformal superalgebra\vspace{-1ex}}
\author{%
    L.A.Yates and P.D.Jarvis\vspace{10pt}\\
    \footnotesize{School of Physical Sciences, University of Tasmania, Hobart, Tasmania, Australia}\\
    \footnotesize{Email: Luke.Yates@utas.edu.au and Peter.Jarvis@utas.edu.au}\\
    }
\date{}
\maketitle

\abstract
We analyze the structure of the family of quadratic superalgebras, introduced in J Phys A 44(23):235205 (2011), for the quadratic deformations of $N = 1$ space-time conformal supersymmetry. We characterize in particular the `zero-step' modules for this case. In such modules, the odd generators vanish identically, and the quadratic superalgebra is realized on a single irreducible representation of the even subalgebra (which is a Lie algebra). In the case under study, the quadratic deformations of $N = 1$ space-time conformal supersymmetry, it is shown that each massless positive energy unitary irreducible representation (in the standard classification of Mack), forms such a zero-step module, for an appropriate parameter choice amongst the quadratic family (with vanishing central charge). For these massless particle multiplets therefore, quadratic supersymmetry is unbroken, in that the supersymmetry generators annihilate \emph{all} physical states (including the vacuum state), while at the same time, superpartners do not exist.\\\vspace{10mm}


\newtheorem{theorem}{Theorem}[section]
\newtheorem{defn}[theorem]{Definition}
\newtheorem{example}[theorem]{Example}
\newtheorem{remark}[theorem]{Remark}
\newtheorem{lemma}[theorem]{Lemma}


\section{Introduction}
\label{sec:Introduction}
While supersymmetry in physics can trace its
origins to relativistic field quantization and the spin-statistics theorem
itself, its emergence as a symmetry principle governing  
transformations amongst field multiplets has come to dominate
thinking in the era of unification and model building 
based upon the gauge principle for matter-force
interactions. Important axiomatic milestones are the realization
that supersymmetry is uniquely able to circumvent
otherwise proscriptive `no-go' theorems \cite{Coleman&Mandula1967,ORaifeartaigh1975} on 
combining internal and space-time symmetry, matched
by a restricted set of possibilities for
its algebraic stucture under the strictures of
local relativistic quantum fields in four dimensions \cite{HLS1975}. 
Theoretically its appeal stems from the improved behaviour
of supersymmetric quantum field theories
with respect to infinities and renormalization \cite{Sohnius1985,West1990}.

Despite several decades of experiment, including
recent LHC high energy searches, signals of 
(broken) supersymmetry have however remained absent from 
the particle data (see for example
chapter 10 of \cite{Bechtle2015a}). In this work we propose a radical alternative possibility: 
supersymmetry is indeed present (and unbroken), but is 
invisible in the spectrum of physical states.

Given the well-established picture of the
supersymmetric phenomenology of relativistic fields, 
our development necessarily invokes assumptions 
which go beyond the conventional framework \cite{HLS1975}.
At the same time, however, our formulation 
is algebraic, and so robust to detailed
model implementations. 

We posit that the algebra of supersymmetry generators be extended, from a graded Lie structure, to a more flexible
associative superalgebra wherein terms quadratic in the generators of the even part can occur in the anticommutator
of odd generators (a specific type of `quadratic superalgebra' -- definitions will be given below). It is then possible to find `zero-step' modules, wherein the supersymmetry generators are identically zero -- the quadratic superalgebra is carried entirely on a single (irreducible) multiplet of the even subalgebra, thus obviating the need for supersymmetric partners.

The case taken up in the present work is that of the 
conformal symmetry group, $SO(4,2)\cong SU(2,2)$\,, acting on four-dimensional
Minkowski space, its corresponding spacetime Lie superalgebra (corresponding to $N=1$ superconformal symmetry),
and its admissible extensions to a quadratic superalgebra. 
Despite quite rigid algebraic constraints, we show that the class of 
unitary zero step modules coincides with that of the 
lowest energy unitary irreducible massless representations
of the conformal group, in the classification of Mack \cite{Mack77}. 
Thus, remarkably, the known particle states of the standard model 
(all spin-$\frac 12$ matter fields, spin-$1$
gauge bosons and Higgs scalars), whose
masses are indeed negligible relative to unification scales, 
 are descended from conformal massless 
multiplets at high energy, which are in fact zero-step multiplets 
of the quadratic conformal superalgebra.

It should be noted that the way in which 
`hidden supersymmetry' arises in our present development differs from
that of other contexts that have been noted in the literature.
There are many works analyzing the symmetries
of certain quantum (and some stochastic) systems where 
the structures and state spaces are naturally graded, and 
so admit hitherto unidentified odd generators, to play the role of 
supersymmetry charges. This applies particularly to 
quantum mechanical problems \cite{CooperFreedman1983}, analyzed using the factorization
method \cite{Infeld1951}, where supersymmetry provides new insights.
Also different is a class of supergravity field theories
entailing super-transformations on field multiplets which are not linear 
representations, deemed to manifest `secret supersymmetry'  
\cite{SamuelWess1984}. The name derives
from the fact that, in comparison with standard linear actions
there is a relative paucity of super-partners
(but not, as in our case, their complete absence).

The aim of the present paper is to consolidate and elaborate both the theoretical background, and physical context, of the results first adumbrated in \cite{Jarvis2012}, an application to spacetime supersymmetry of the analysis of 
quadratic superalgebras given in \cite{JarvisRudolphYates2011}.  In \S \ref{sec:QuadraticSuperalgebras} below, we reiterate the main results of \cite{JarvisRudolphYates2011}, in the context of  the consistent formulation of generator relations and constraints. We are at pains to establish the Poincar\'{e}-Birkhoff-Witt (PBW) property of the enveloping algebra, for which we provide a new proof, using the broader theory of quadratic algebras (references will be given below). This applies in particular to the case of the quadratic superalgebras $gl_2(n/1)$\, the main focus of our work. 

In \S \ref{sec:QSuperconformalAlgebra}, we turn to the $n=4$ case, and specifically its real form $su_2(2,2/1)$\,, which admits as a contraction limit the algebra of $N=1$ conformal supersymmetry in four dimensions, $SU(2,2/1)$  (standard notation and results for $su(2,2/1)$ itself, including the classification due to Mack \cite{Mack77} of the massless unitary irreducible modules of $su(2,2)$, are summarized in  \S \ref{sec:superconformalAlgebra} ). %
%
Exploiting the so-called degenerate nature of these massless unirreps and using the general theory of characteristic polynomial identities for Lie algebras (for which necessary results are outlined), we show that each of these representations admits a minimal quadratic polynomial identity which brings it into coincidence with the odd super-charge generator anticommutator bracket of one member (that is, for a specific parameter choice) of the $su_2(2,2/1)$ quadratic superalgebra class, thus confirming \cite{Jarvis2012}. \S \ref{sec:Conclusion} concludes with further observations on the unbroken supersymmetry without super-partner scenario, and on the mathematical level, on quadratic superalgbras as a feasible `nonlinear' extension of supersymmetry principles. 

The paper contains
several appendices: 
In \S \ref{subsec:Isomorphisms} we analyze the full parametric form of admissible
anticommutator structure constants for $gl_2(n/1)$\,, and identify the isomorphic classes in parameter space, which 
arise via rescaling and shifting on certain generators. For completeness we also include variants wherein the odd generators transform as densities rather than vector operators under $gl(n)$, a case in point being an explicit fermionic oscillator construction for $n=4$\,. \S \ref{subsec:PBWconditions} includes an independent proof of the important Poincar\'{e}-Birkhoff-Witt (PBW) property of the enveloping algebra of the class of quadratic superalgebras under consideration, as an application of the so-called Diamond Lemma (for references see text). In particular, it is shown that the inclusion of additional quadratic relations of odd type, which for example arise naturally in the fermionic realization, lead to enveloping algebras which 
are not of PBW type.  Finally \S \ref{subsec:MasslessnessAndBosonicOscillatorModel} on $su(2,2)$\, supplements the overview given in  \S \ref{sec:QSuperconformalAlgebra} by re-deriving the necessary and sufficient conditions for massless multiplets of \cite{Mack77} and also provides a review of concrete Fock space constructions of 
these (that is, in an algebra realized via boson creation and annihilation operators).


\section{Quadratic Superalgebras}
\label{sec:QuadraticSuperalgebras}
We begin with a review of the class of quadratic deformations of Lie superalgebras introduced in \cite{Jarvis&Rudolph2003} and \cite{JarvisRudolphYates2011}. %

Let $L=L_{\bar{0}}+L_{\bar{1}}$ be a finite-dimensional $\mathbb{Z}_2$-graded complex vector space. Take the even and odd subspaces $L_{\bar{0}}$ and $L_{\bar{1}}$ to be spanned by basis elements $x_1,\,i=1,2,...,n$, and $y_r,\,r=1,2,...,m$, respectively. The tensor algebra $T(L)=\sum_{n=0}^{\infty}\otimes^n(L)\cong\mathbb{C}+ L+L\otimes L+\cdots$ inherits the $\mathbb{Z}_2$-grading in the natural way, in that for $T^n:=\otimes^n(L)$ we have
\begin{align}
\left. T^n \right|_{\bar{0}} \cong  & \,  {\sum}_{\sum \bar{\alpha}_i \equiv \bar{0}} L_{\bar{\alpha}_1} \otimes L_{\bar{\alpha}_2} \otimes \cdots  \otimes L_{\bar{\alpha}_n}, \qquad
\left. T^n \right|_{\bar{1}} \cong  {\sum}_{\sum \bar{\alpha}_i \equiv \bar{1}}  L_{\bar{\alpha}_1} \otimes L_{\bar{\alpha}_2} \otimes \cdots  \otimes L_{\bar{\alpha}_n}, \nonumber
\end{align}
with each $\bar{\alpha}_i =\bar{0}$ or $\bar{1}$, and $T^n \cong \left. T^n \right|_{\bar{0}}+\left. T^n \right |_{\bar{1}}$.
We impose the structural relations
\begin{align*}
	[L_{\bar0},L_{\bar0}]\subset L_{\bar0},\quad[L_{\bar0},L_{\bar1}]\subset L_{\bar1}\quad\mbox{and}\quad
	\{L_{\bar1},L_{\bar1}\}\subset L_{\bar0}\otimes L_{\bar0} + L_{\bar0} + \mathbb{C},
\end{align*}
which in terms of basis elements take the form
\begin{align}\label{eq:qlsa:commRelations}
	[x_i,x_j]=c_{ij}{}^kx_k\qquad [x_i,y_p]=\bar{c}_{ip}{}^qy_q\qquad \{y_p,y_q\}=d_{pq}{}^{kl}x_k\otimes x_l + b_{pq}{}^kx_k + a_{pq},
\end{align}
where the arrays of complex numbers $c_{ij}{}^k$, $\bar{c}_{ip}{}^q$, $b_{pq}{}^k$ and $d_{pq}{}^{kl}$ take the role of generalised structure constants. Demanding that $[\phantom{A},\phantom{A}]$ and $\{\phantom{A},\phantom{A}\}$ fulfill the standard (graded) commutation relations the relations \eqref{eq:qlsa:commRelations} define a set of quadratic relations $I\subset L\otimes L + L + \mathbb{C}$ spanned by the set
\begin{align}\label{eq:qlsa:quadraticRelations}\begin{array}{l}
	x_i\otimes x_j - x_j\otimes x_i - c_{ij}{}^kx_k;\\
	x_i\otimes y_p - y_p\otimes x_i - \bar{c}_{ip}{}^qy_q;\\
	y_p\otimes y_q - y_q\otimes y_p -d_{pq}{}^{kl}x_k\otimes x_l - b_{pq}{}^kx_k -a_{pq}.\end{array} 
\end{align}
Next we demand that the relations \eqref{eq:qlsa:commRelations} satisfy the graded Jacobi identities,
\begin{align*}
	[x,[y,z]]=[[x,y],z]+(-1)^{|x|.|y|}[y,[x,z]]]
\end{align*}
for homogneous $x,y,z\in L$ with $|x|,|y|=\bar0$ or $\bar1$ being the $\mathbb{Z}_2$-grading of $x$ and $y$ respectively. In terms of the generalised structure constants the Jacobi identities take the form
\begin{align}\label{eq:qlsa:jacobiIdentities}\begin{array}{l}
{c_{ij}}^l{c_{l k}}^m= \, {c_{ik}}^l{c_{l j}}^m+{c_{jk}}^l{c_{l i}}^m ;\\
{c_{ij}}^l \bar{c}_{l p}{}^q =  \, \bar{c}_{ir}{}^q\bar{c}_{jp}{}^r-\bar{c}_{jr}{}^q\bar{c}_{ip}{}^r ;\\
{c_{in}}^k {d_{pq}}^{nl}+ {c_{in}}^l {d_{pq}}^{k n} =  \,
\bar{c}_{ip}{}^s {d_{sq}}^{kl} + \bar{c}_{ip}{}^s {d_{sq}}^{kl} ,\\
{b_{pq}}^m {c_{im}}^n =  \, \bar{c}_{ip}{}^s {b_{sq}}^n + \bar{c}_{iq}{}^s {b_{ps}}^n;\\
\bar{c}_{mp}{}^s {b_{qr}}^m + \bar{c}_{mq}{}^s {b_{rp}}^m + \bar{c}_{mr}{}^s {b_{pq}}^m =  \, 0,\\
\bar{c}_{mp}{}^s {d_{qr}}^{ml} + \bar{c}_{mq}{}^s {d_{rp}}^{ml} + \bar{c}_{mr}{}^s {d_{pq}}^{ml} =  \, 0.
\end{array}
\end{align}
\begin{defn}[Quadratic Superalgebra]
Let $L=L_{\bar{0}}+L_{\bar{1}}$ be a finite-dimensional $\mathbb{Z}_2$-graded complex vector space satisfying the  multiplicative relations \eqref{eq:qlsa:commRelations} and the Jacobi identities \eqref{eq:qlsa:jacobiIdentities} above. Let $(I)$ be the two-sided ideal in the tensor algebra $T(L)$ generated by the set of qudratic relations $I$ as in \eqref{eq:qlsa:quadraticRelations}. The subalgebra of the tensor algebra defined by $U(L)=T(L)/(I)$ is called the quadratic superalgebra associated with $I$.\end{defn} 

Note when $d_{pq}{}^{kl}=0$ then $U(L)=T(L)/(I)$ is the universal enveloping algebra of the ordinary Lie superalgebra $L=L_{\bar{0}}+L_{\bar{1}}$. In analogy with the ordinary Lie superalgebra case we define type I quadratic superalgebras to be those such that $L_{\bar1}=L_+ + L_-\cong V_0(\lambda)+V_0(\lambda^*)$  where  $V_0(\lambda)$ is a (complex) irreducible $L_{\bar0}$-module with highest weight $\lambda$ and $V_0(\lambda^*)$ its contragredient. In the absence of classification theorems, we also impose $[L_{\pm},L_{\pm}]=0$ by analogy with the type I Lie superalgebra case. Type II quadratic superalgebras have the structure $L_{\bar1}\cong V(\lambda)$ for $V_0(\lambda)$ a real representation. 

It was shown in \cite{JarvisRudolphYates2011} that by employing various oscillator constructions several examples of type I quadratic superalgebras may be constructed. In each of these cases the even part is isomorphic to $gl(n)$ and the odd part is characterised by $\lambda$ equal to one of $\{1\}$, $\{1,1,1\}$ and $\{3\}$ (in partition notation). Also discovered was a parametrised class of type I quadratic superalgebras which do not rely on concrete oscillator constructions. This class permits, in the case of the even subalgebra, a presentation in terms of the abstract Gel'fand generators of $gl(n)$. The odd part consists of the standard weight vectors associated with the corresponding fundamental representation and its contragredient.\\

\noindent$\mathbf{L=gl_2(n/1)}\vspace{2mm}\\
$ Let $L_{\bar0}=gl(n)$ and $L_{\bar1}=L_{-1}+L_{+1}=\{1\}+\{\overline{1}\}$.
The even generators are $E^a{}_b$ satisfying the abstract relations
\begin{align}\label{eq:gelfand_relations}
[E^a{}_b,E^c{}_d]=\delta^c{}_bE^a{}_d-\delta^a{}_dE^c{}_b.\qquad a,b,c,d=1,..,n	
\end{align}
The odd generators are denoted $\overline{Q}^a$ and $Q_a$ for $L_{+1}$ and $L_{-1}$ respectively. Under the adjoint action of $gl(n)$ these transform as vector operators satisfying
\begin{align}\label{eq:adActionOdd}
	[E^a{}_b,\overline{Q}^c]=\delta^c{}_b\overline{Q}^a\qquad [E^a{}_b,Q_c]=-\delta^a{}_cQ_b.
\end{align}
We impose $\{\overline{Q}^a,\overline{Q}^b\}=0=\{Q_a,Q_b\}$ and finally the quadratic bracket relations between $\overline{Q}$ and $Q$ read
\begin{align}
\label{eq:generic_solution}
{\{}\overline{Q}{}^a, Q_b{\}} = 
(E^2){}^a{}_b - E{}^a{}_b\big(\langle E \rangle - \alpha\big)
-&\,\textstyle{\frac 12}\delta{}^a{}_b\big(\langle E^2 \rangle \!-\! \langle E \rangle^2 \!+\! (n\!-\!1\!+2\alpha)\langle E \rangle \big) 
+ {\texttt c}\delta^a{}_b\,.
\end{align}
where $(E^2)^a{}_b=E^a{}_cE^c{}_b$, $\langle E\rangle:=\sum_iE^i{}_i$, $\langle E^2\rangle=\sum_i(E^2)^i{}_i$ are the linear and quadratic Casimirs, and ${\sf c}$ is a (complex) central term. Note the linear Casimir satisfies $[\langle E\rangle,\overline{Q}^a] =\overline{Q}^a$ and $[\langle E\rangle,Q_a]={-Q_a}$. This results in a one-parameter family $gl_2(n/1)=gl_2(n/1)^{\alpha,\,{\sf c}}$ of quadratic superalgebras plus a central charge.\\

Parametrised solutions for the anticommutator are derived by taking the most general form of the anticommuator compatible with $gl(n)$ covariance and imposing the Jacobi identity \eqref{eq:qlsa:jacobiIdentities}. The $gl(n)$-compatible form is recalled in \S \ref{subsec:Isomorphisms} and the specific instance  \eqref{eq:generic_solution} corresponds to ${\sf a}\rightarrow 1$, which suppresses an overall normalisation, and ${\sf b}_2\rightarrow \alpha$ (compare \eqref{eq:anticomm_gln_covariant}). The close relationship between $gl_2(n/1)$ and Lie superalgebra $sl(n/1)$ is revealed by a simple re-parametrisation. Let $gl_2(n/1)^{\lambda,\,{\sf c}}$ denote the quadratic superalgebra resulting from ${\sf a}\rightarrow \lambda\equiv\frac{1}{\alpha}$ and ${\sf b}_2\rightarrow\lambda\alpha=1$, this is equivalent to an overall rescaling of \eqref{eq:generic_solution} by $\lambda$, explicitly  
\begin{align*}
{\{}\overline{Q}{}^a, Q_b{\}} =E{}^a{}_b-\delta^a{}_b\langle E \rangle+\lambda\big[(E^2){}^a{}_b - E{}^a{}_b\langle E \rangle -&\,\textstyle{\frac 12}\delta{}^a{}_b\big(\langle E^2 \rangle \!-\! \langle E \rangle^2 \!+\!(n\!-\!1)\langle E \rangle -2{\sf c} \big)\big].
\end{align*}

\begin{lemma}[Contraction limit of $gl_2(n/1)^{\lambda,\,{\sf c}}$]\label{lemma:contraction_limit}
The ordinary Lie superalgebra $sl(n/1)$ is obtained from $gl_2(n/1)^{\lambda,\,{\sf c}}$ in the contraction limit $\lambda\rightarrow0$.
\end{lemma}

%

The class of quadratic superalgebras can be placed within the much larger class of quadratic algebras. In this setting there exists a generalisation of the classical PBW theorem to which there corresponds a set of generalised Jacobi identities and in certain cases an algorithm to derive a basis of ordered monomials. We provide below a brief account of the main definitions and theorems following closely the text of Polishchuk and Positselski \cite{Polishchuk2005} and key references therein such as Braverman \& Gaitsgory \cite{Braverman1996} and Priddy \cite{Priddy1970}. These results are used to derive an alternative proof of the PBW Theorem for quadratic superalgberas, see Theorem \ref{thm:PBW_QLSA}.

Let $X$ be a finite-dimensional vector space of dimension $n$ and let $T(X)$ be the tensor algebra generated by $X$. We fix a set of \emph{non-homogeneous quadratic relations} $I\subset (T\otimes T) \oplus T \oplus \mathbb{C}$ to which there corresponds a set of \emph{homogeneous relations} $I_2\subset T\otimes T$ which are the projection of $I$ onto $T\otimes T$. We denote by $(I)$ and $(I_2)$ the ideal generated in $T(X)$ by $I$ and $I_2$ respectively. 
\begin{defn}[Quadratic Algebra]
	Let $X$ and $I$ be defined as above. The algebras
	\begin{align*}
		U=T(X)/(I)\qquad\mbox{and}\qquad A=T(X)/(I_2).
	\end{align*}
	are called the \textbf{inhomogeneous quadratic algebra} and the \textbf{homogenous quadratic algebra} respectively, generated by $X$ and $I$.
\end{defn}
Note that the direct sum decomposition $I\subset X\otimes X + X+\mathbb{C}$ enables maps $\alpha:I_2\rightarrow X$ and $\beta:I_2\rightarrow\mathbb{C}$ to be defined such that
\[I=\{x-\alpha(x)-\beta(x)|x\in I_2\}.\]
Within the class of quadratic algebras there exists an important subclass possessing certain cohomological properties which permit a generalisation of the classical PBW theorem. This class contains only homogenous quadratic algebras and is defined by the notion of Koszulness for which there are many equivalent definitions. We give the following definition in terms of the distributivity of certain vector subspaces in the tensor algebra:
\begin{defn}[Koszul Algebra(\cite{Polishchuk2005}, chapter 2, theorem 4.1)] A homogeneous quadratic algebra $A = T(X)/(I_2)$ is Koszul iff for all $n\geq0$ the collection of subspaces
\[X^{\otimes i-1}\otimes I_2 \otimes X^{\otimes n-i-1}\subset X^{\otimes n},\quad i=1,...,n-1\]
is distributive.
\end{defn}
As a final preliminary to the generalised PBW theorem we recall the following. Associated with the tensor algebra is the filtration defined by $T_n=\sum_{k=0}^n T^k$ in such a way that $\mathbb{C}\cong T_0\subset T_1 \subset T_2 \subset \cdots$, that is, $T_n\subset T_{n+1}$. $U$ inherits this filtration in the natural way so that we have $\mathbb{C}\cong U_0\subset U_1 \subset U_2 \subset \cdots$ and we define the \emph{associated graded algebra} as the direct sum 
\begin{align}
	grU\equiv \bigoplus_n^{\infty} U_n/U_{n-1}.
\end{align}
Let us compare the construction of $A$ with that of $grU$\footnote{For a more detailed discussion along these lines we refer the reader to the review \cite{Shepler2014}}. $A$, the homogeneous version of $U$, is generated by first homogenising, that is truncating, each term in the generating relations and then factoring the tensor algebra by the resulting ideal. The construction of $grU$, on the other hand, is obtained by initially retaining the full set of non-homogeneous relations and instead truncating the terms appearing in the corresponding ideal. The generalisation of the PBW theorem is a statement of the conditions under which these two graded algebras coincide.

\begin{theorem}[Generalised PBW theorem(\cite{Polishchuk2005}, chapter 5, theorem 2.1)]\label{thm:gen_PBW} When $A$ is Koszul, and the following conditions are satisfied:
\begin{align}\label{eq:generalisedJacobi}
	\begin{array}{ll}
	\mbox{(J1)}\quad & (\alpha\otimes {\sf id}-{\sf id}\otimes\alpha)|_{I_2\otimes L\cap L\otimes I_2}\subset I_2;\\
	\mbox{(J2)} & \alpha\circ(\alpha\otimes {\sf id}-{\sf id}\otimes\alpha)|_{I_2\otimes L\cap L\otimes I_2}
	=-(\beta\otimes {\sf id}-{\sf id}\otimes\beta)|_{I_2\otimes L\cap L\otimes I_2}\\
	\mbox{(J3)} & \beta\circ(\alpha\otimes {\sf id}-{\sf id}\otimes\alpha)|_{I_2\otimes L\cap L\otimes I_2}=0
	\end{array}
\end{align}
we have the isomorphism
\[gr(U)\cong A.\]
\end{theorem}
The relations (J1)-(J3) are called the \emph{generalised Jacobi identities}. Proving Koszulness is in general a difficult undertaking and one may select from a variety of methods (for example \cite{Polishchuk2005}, chapter 2). Here instead we investigate when the homogeneous algebra $A$ of theorem \ref{thm:gen_PBW} satisfies the stronger condition of being a \emph{PBW algebra}, that is, admitting an ordered basis of monomials in the generators $x_i\in X$ (see definition \ref{defn:PBW_Algebra}). A theorem due to Priddy (\cite{Priddy1970}, theorem 5.3) states that every homogeneous PBW algebra is Koszul.

Before giving a formal definition of a PBW algebra we must make precise the notion of an ordered set of monomials with respect to a set of quadratic relations. Following \cite{Priddy1970} let $x_i$, $i\in S_1 := \{1,..,n\}$ be a basis for $X$. $S_1$ defines an ordering on $X$ such $x_i<x_j$ when $i<j$. $S_1$ also defines a lexicographic ordering on monomials belonging to $T(X)$. $I_2$ comprises expressions of the form $\sum{c^{kl}x_kx_l}$. Each such expression contains a unique \emph{leading monomial} which is the highest quadratic term, with respect to the lexicographic ordering, in the sum. We define
\begin{align*}
	S_2 := \{(i,j)|i,j\in S_1,\,x_ix_j \mbox{ is not a leading monomial}\}.
\end{align*}
We also define 
\begin{align*}
	S_i:=\{(i_1,i_2,...,i_n)|(i_{j},i_{j+1})\in S_2, j=1,2,...,n-1\}.
\end{align*}
Consider now the grading of $A$ inherited from $T(X)$. Since $I_2$ is by construction homogeneous each graded subspace $A_i$ contains only homogeneous elements of degree $i$. It follows that
\begin{align*}
	&A_0\cong \mathbb{C},\\
	&A_1\cong X \mbox{ has a basis } \{x_i|x_i\in X,\,i\in S_1\},\\
	&A_2 \mbox{ has a basis } \{x_ix_j|x_i,x_j\in X,\,(i,j)\in S_2\}.
\end{align*}

\begin{defn}\label{defn:PBW_Algebra}
The homogenous quadratic algebra $A$ is a \emph{PBW-algebra} if the monomials $(x_{i_1}x_{i_2}...x_{i_n}$, $(i_1,i_2,...,i_n)\in S_n)$ are a basis for $A$. In this case the monomials are called a \emph{PBW-basis} of $A$.
\end{defn}
Note that these monomials always span $A$ thus the task of establishing the PBW property is to determine their linear independence. In order to proceed with this task we define the mapping $\pi:X\otimes X \rightarrow X\otimes X$ 
\begin{equation*}
 \pi(x_i,x_j) = \left \{
\begin{array}{ll}
	x_ix_j & (i,j)\in S_2 \\ 
	\sum_{(k,l)\in S_2}{c^{kl}x_kx_l}\quad &(i,j)\notin S_2
\end{array}
\right.
\end{equation*}
which extends linearly to $X\otimes X$ and essentially replaces leading monomials with a unique sum of non-leading terms as determined by $I_2$. Finally we define
\begin{align*}
	\pi^{12}=\pi\otimes I:T_3\rightarrow T_3.\qquad
	\pi^{23}=I\otimes\pi:T_3\rightarrow T_3.
\end{align*}
\begin{lemma}[Thm 2.1 p.82 \cite{Polishchuk2005} - Diamond Lemma]\label{lemma:diamond}
$A$ is a PBW-algebra iff the cubic monomials ($x_ix_jx_k$, $(i,j,k)\in S_3$) are linearly independent in $A_3$. Equivalently $A$ is a PBW-algebra iff the following equation holds:
\begin{align}\label{eq:diamondLemma}
	\cdots\pi^{12}\pi^{23}\pi^{12}\pi^{23}\pi^{12} = \cdots\pi^{23}\pi^{12}\pi^{23}\pi^{12}\pi^{23}. 
\end{align}
\end{lemma}
\textbf{Remarks.} The infinite composition is well defined since $\pi$ decreases the order. To establish \eqref{eq:diamondLemma} we need only consider basis elements $x_ix_jx_k\in T_3$ such that both $(i,j),(j,k)\notin S_2$. For if one of these belonged to $S_2$ then one of either $\pi^{12}$ or $\pi^{23}$ will act trivially on the starting term $x_ix_jx_k$ and \eqref{eq:diamondLemma} follows immediately. The PBW property depends on the choice of ordering given to the generators of $X$. That is a fixed homogeneous quadratic algebra may be a PBW algebra given one ordering but not for another. 
\begin{lemma}[Lemma 2.9 \cite{JarvisRudolphYates2011}]\label{lemma:QSLAisPBW}A homogeneous quadratic superalgebra	 is a PBW-algebra under any index ordering such that only those $d_{pq}{}^{kl}$ are nonvanishing for which $k$, $l$ precedes $p$, $q$. 
\end{lemma}
\textbf{Alternative Proof.} Using Lemma \ref{lemma:diamond}. See appendix \ref{subsec:PBWconditions}.

\begin{lemma}[Jacobi identities for QLSA]\label{lemma:JacobiEquivalence} The generalised Jacobi identities \eqref{eq:generalisedJacobi} are equivalent to \eqref{eq:qlsa:jacobiIdentities}.
\end{lemma}
\textbf{Proof.} See Lemma 2.8 \cite{JarvisRudolphYates2011}.\qed\\

\begin{lemma}[PBW Theorem for Quadratic Superalgebras.] \label{thm:PBW_QLSA} Under the ordering conditions of Lemma \ref{lemma:QSLAisPBW} a quadratic superalgebra has a PBW basis of ordered monomials.
\end{lemma}
\textbf{Proof.} The required monomials are inherited from the homogeneous algebra as per Theorem \ref{thm:gen_PBW}   and due to Lemmas \ref{lemma:QSLAisPBW} and \ref{lemma:JacobiEquivalence}.\qed


\section{Superconformal Algebra}\label{sec:superconformalAlgebra}
In this section we turn to the algebra of the superconformal spacetime symmetry group which is isomorphic to the Lie superalgebra $su(2,2/1)$. The even part is the real form $L_{\bar0}\cong u(2,2)\cong su(2,2)+gl(1)$ and the odd part is the $L_0$-module $L_{\bar1}\cong\{\overline{1}\}+\{1\}$. In addition to the basis of Gel'fand generators and vector operators, we introduce in this section several alternative bases and establish the mappings between them and other bases appearing in the literature. We recall the classification of all positive energy unitary representations of the even subalgebra as originally given by Mack \cite{Mack77}. In particular we examine the class of massless representations for which we provide in \S \ref{subsec:Isomorphisms} a re-derivation of the massless conditions in our preferred basis. Finally, within the broader classification of unitary representations of $su(2,2)$ as given by Yau \cite{Yao1968}, we note that all massless representations have the property of being so-called degenerate representations.

As a basis for $su(2,2/1)$ we choose for the even part the Gel'fand generators $E^a{}_b$, $a,b=1,...,4$ and for the odd part the vector operators $\overline{Q}^c$ and $Q_c$, $c=1,...,4$. The nonzero commutation relations are
\begin{subequations}\label{eq:defRelationsSconformalGelfand}
\begin{align}
&[E^a{}_b,E^c{}_d]=\delta^c{}_bE^a{}_d-\delta^a{}_dE^c{}_b.\qquad a,b,c,d=1,..,n\label{eq:comm_relns_su22_a}\\
&[E^a{}_b,\overline{Q}^c]=\delta^c{}_b\overline{Q}^a\quad [E^a{}_b,Q_c]=-\delta^a{}_cQ_b\label{eq:comm_relns_su22_b}\\
&\{\overline{Q}^c,Q_d\}=E^c{}_d + \delta^c{}_d Z,
\end{align}
\end{subequations}
where 
\begin{align}\label{eq:linear_Casimir_su22}
	Z\equiv -\langle E\rangle=-E^c{}_c
\end{align}
is the linear Casimir. In this non-compact case, the requirement of having unitary representations imposes the following hermiticity conditions
\begin{align}\label{eq:hermiticityConditions}
	(E^a{}_b)^\dagger=\eta^b{}_{b'}E^{b'}{}_{a'}  \eta^{a'}{}_{a},\qquad(Q_a)^\dagger=\eta^a{}_{b'}\overline{Q}^{b'},
\end{align}
where $\eta=$diag$(-1,-1,1,1)$ \footnote{This basis may easily be brought into correspondence with that of \cite{Binegar1986,	Flato1984}. Set $L^a{}_b=E^a{}_b-\frac14\delta^a{}_b\langle E\rangle$ and define $T^a{}_b=\eta^a{}_{a'}L^{a'}_{b'}\eta^{b'}{}_b$ satisfying 
\[[T^a{}_b,T^c{}_d]=\eta^a{}_dT^c{}_b-\eta^c{}_bT^a{}_d.\]
These generators together with $\overline{Q}^a$, $Q_a$ and $Z$ satisfying $[Z,L^a{}_b]=0$, $[Z,\overline{Q}^a]=-\overline{Q}^a$ and $[Z,Q_a]=Q_a$ form the required basis.}. The advantage of this basis is its direct correspondence with the generators of the quadratic superalgebra $gl_2(4/1)$ as well as its straightforward connection with the characteristic identities of $gl(n)$ to be explored in \S \ref{sec:QSuperconformalAlgebra}. 
 
An alternative basis for the even part arises directly from the isomorphism of real Lie algebras $so(4,2)\cong su(2,2)$. We denote by $J_{AB}=-J_{BA}$, $A,B=0,1,2,3,5,6$ the generators of $so(4,2)$. These together with the additional $gl(1)$ generator $Z$ satisfy the commutation relations 
\begin{align}\label{eq:so42_commRels}
\begin{array}{ll}
	&[J_{AB},J_{CD}]={\sf i}(g_{BC}J_{AD}-g_{AC}J_{BD}+g_{AD}J_{BC}-g_{BD}J_{AC})\\
	&[J_{AB},Z]=0
\end{array}
\end{align}
where $g=$diag$(1,-1,-1,-1,-1,1)$. To facilitate a mapping between this basis for $so(4,2)+gl(1)$ and the previous for $u(2,2)$ we define in the fundamental representation the following basis for $u(2,2)$
\begin{equation}
\begin{aligned}\label{eq:concretePrefBasis}
	&m_0=\frac12\left(\begin{array}{cc}\mathbb{I}&0\\0&0\end{array}\right)\quad
	n_0=\frac12\left(\begin{array}{cc}0&0\\0&\mathbb{I}\end{array}\right)\quad
	x^+_0=\frac12\left(\begin{array}{cc}0&\mathbb{I}\\0&0\end{array}\right)\quad
	x^-_0=\frac12\left(\begin{array}{cc}0&0\\ \mathbb{I}&0\end{array}\right)\quad\\
	&m_i=\frac12\left(\begin{array}{cc}\sigma_i&0\\0&0\end{array}\right)\quad
	n_i=\frac12\left(\begin{array}{cc}0&0\\0&\sigma_i\end{array}\right)\quad
	x^+_i=\frac12\left(\begin{array}{cc}0&\sigma_i\\0&0\end{array}\right)\quad
	x^-_i=\frac12\left(\begin{array}{cc}0&0\\ \sigma_i&0\end{array}\right)\quad
\end{aligned}
\end{equation}
where $\sigma_i$, $i=1,2,3$ are the standard Pauli matrices and $\mathbb{I}=$diag$(1,1)$. The corresponding abstract generators are obtained by replacing elementary $e^i{}_j$ with $E^i{}_j$, in particular we have
\begin{equation}
\begin{aligned}
		&M_i=\frac12(e^1{}_1\otimes\sigma_i)^{{\sf T}\,b}{}_a E^a{}_b\qquad\quad
		&N_i=\frac12(e^2{}_2\otimes\sigma_i)^{{\sf T}\,b}{}_a E^a{}_b\\
		&X^+_i=\frac12(e^1{}_2\otimes\sigma_i)^{{\sf T}\,b}{}_a E^a{}_b
		&X^-_i=\frac12(e^2{}_1\otimes\sigma_i)^{{\sf T}\,b}{}_a E^a{}_b.
\end{aligned}
\end{equation}
These satisfy the commutation relations
\begin{align}\label{eq:comm_relations_preferred}
	\begin{array}{ll}
		{}[M_i,M_j]={\sf i}\varepsilon_{ijk}M_k \qquad\qquad  &[N_i,N_j]={\sf i}\varepsilon_{ijk}N_k\\
		{}[M_0,X^+_0]=\frac12 X^+_0\qquad & [N_0,X^+_0]=-\frac12 X^+_0\\
		{}[M_0,X^+_i]=[M_i,X^+_0]=\frac12 X^+_i\qquad & [N_0,X^+_i]=[N_i,X^+_0]=-\frac12 X^+_i\\
		{}[M_i,X^+_j]=\frac12 i\varepsilon_{ijk}X^+_k + \delta_{ij}X^+_0 & [N_i,X^+_j]=\frac12 i\varepsilon_{ijk}X^+_k-\delta_{ij}X^+_0\\
		{}[X^+_0,X^-_0]=\frac12 (M_0-N_0) & [X^+_i,X^-_j]=\delta_{ij}\frac12 (M_0-N_0) 
	\end{array}
\end{align}
Following Wybourne \cite{Wybourne} (see also Appendix C \cite{Stoyanov1968}) we introduce an extended set of Dirac matrices,
\begin{align*}
	\gamma_0=\sigma_3\otimes\mathbb{I}=\left(\begin{array}{cc}\mathbb{I}&0\\0&-\mathbb{I}\end{array}\right)\qquad
	\gamma_i={\sf i}\sigma_2\otimes\sigma_i=\left(\begin{array}{cc}0&\sigma_i\\-\sigma_i&0\end{array}\right)\qquad
	\gamma_5\equiv \gamma_0\gamma_1\gamma_2\gamma_3=\left(\begin{array}{cc}0&-{\sf i}\\-{\sf i}&0\end{array}\right)
\end{align*}
and define the 6-component vector $\hat{\gamma}_A=\left(-\gamma_1,-\gamma_2,-\gamma_3,\gamma_5,\gamma_0,-{\sf i}\mathbb{I}\right).$
The operators
\begin{align}\label{eq:gammaRepSO42}
	j_{AB} = \frac{{\sf i}}{2}\hat{\gamma}_A\hat{\gamma}_B\qquad A<B,\quad A,B = 1,2,3,4\equiv0,5,6.
\end{align}
together with $z=\frac12(\sigma_0\otimes\mathbb{I})$ provide a four-dimensional matrix representation of $so(4,2)+gl(1)$ satisfying \eqref{eq:so42_commRels}. Explicit calculation of \eqref{eq:gammaRepSO42} combined with \eqref{eq:concretePrefBasis} generates the following map from the abstract generators of $so(4,2)+gl(1)$ to those of $u(2,2)$:  
\begin{align}\label{eq:so42ToPref}
\begin{array}{llll}
	J_{ij} &= \varepsilon^{ijk}(M_k+N_k)\qquad
	&J_{i5} &= -(M_i-N_i)\\
	J_{45} &= X^+_0-X^-_0\qquad
	&J_{i4} &= {\sf i}(X^+_i+X^-_i)\\
	J_{i6} &= -(X^+_i-X^-_i)\qquad
	&J_{56} &= {\sf i}(X^+_0-X^-_0)\\
	J_{46} &= M_0-N_0\qquad
	&Z &= M_0+N_0.
	\end{array}
\end{align}
The advantage of the $J_{AB}$ basis is its direct connection with the physical generators of translations, conformal and Lorentz transformations and dilatation, see \eqref{eq:conformal_generators}. The basis of $M$, $N$, $X^+$, $X^-$ acts an intermediary between the $J_{AB}$ and the $E^i{}_j$ bases which facilitates computations and the comparability of results. 

For $sl(4)$ and its real forms, namely $su(2,2)\cong so(4,2)$, the \emph{lowest weight} of a unitary irreducible representation (when it exists) may be characterised by three real numbers. These numbers may be chosen to be the eigenvalues of the Cartan elements under the module action on the lowest weight vector $|\mu\rangle$ and as such they depend on the choice of basis of the Cartan subalgebra.

The same applies for the Lie superalgebra $su(2/2,1)$ where representations are induced from a single representation of the even subalgebra; although in this case an extra label is required owing to the additional Cartan element, the generator of $gl(1)$\footnote{In the massless case, see \eqref{eq:masslessConditionsMack} , representations of the superconformal algebra comprise only $L_{\overline0}$-representations which are themselves massless. This leads to so-called short-multiplets where, irrespective of atypicality, those even submodules which are not massless are decoupled from the induced module, see for example \cite{Binegar1986}\cite{Flato1984}.}. For the remainder of this section we focus exclusively on the representation theory of the subalgebra $su(2,2)\subset L_{\overline0}$ from which representations of $su(2/2,1)$ may be induced.

 We introduce the following Cartan elements for $su(2,2)$ \cite{Mack77}:
\begin{align}\label{basisMack}
\begin{array}{lll}
&H_0\equiv\frac{1}{2}(E^1{}_1+E^2{}_2-E^3{}_3-E^4{}_4) &=M_0-N_0\\ 
&H_1\equiv\frac{1}{2}(E^1{}_1-E^2{}_2) &=M_3\\
&H_2\equiv\frac{1}{2}(E^3{}_3-E^4{}_4) &=N_3.
\end{array}
\end{align}
These action of these on the lowest weight vector $|\mu\rangle$ is
\begin{align}
	H_0|\mu\rangle=d|\mu\rangle,\qquad H_1|\mu\rangle=-j_1|\mu\rangle,\qquad H_2|\mu\rangle=-j_2|\mu\rangle,
\end{align}
where $d\in\mathbb{R}$, known as the conformal dimension, labels reducible representations of the Abelian subgroup generated by $H_0$, and $j_1$ and $j_2$, where $2j_1$ and $2j_2$ are non-negative integers, label representations of the compact subgroup $SU(2)\times SU(2)$ generated by $M_1, M_2, (H_1=M_3)$ and $N_1, N_2, (H_2=N_3)$. Together  these 7 elements generate the \emph{maximal compact subgroup} $SU(2)\times SU(2)\times U(1)$ which plays an key role in the representation theory of $SU(2,2)$; we denote the corresponding subalgebra
\[\mathcal{S}=su(2) \times su(2) \times gl(1).\]
The weight space $H^*$ has a basis of fundamental weights equal in number to the rank $l$, in the present case $l=3$. The fundamental weights may be projected onto an $(l+1)$-dimensional Euclidean space allowing their expression in terms of an orthonormal basis $\varepsilon_i$, $i=1,...,l+1$ satisfying $\varepsilon_i(E^j{}_j)=\delta^i{}_j$ (no sum).
Thus we may write $\mu=\sum_{i=1}^{4}\mu_i\varepsilon_i$ from which it follows
\begin{align}\label{eq:cartesianComponentsMu}
	d=\frac12(\mu_1+\mu_2-\mu_3-\mu_4),\qquad -j_1=\frac12(\mu_1-\mu_2)\qquad \mbox{and} \qquad -j_2=\frac12(\mu_3-\mu_4).
\end{align}
The physically relevant classes of unitary irreducible representations of the conformal group are characterised by their energy, mass and spin/helicity. We write the standard physical basis for the conformal algebra as follows\footnote{This basis appears in the classic texts \cite{Wybourne,West1990,Sohnius1985}. See also \cite{Bracken&Jessup:1982a}.}:
\begin{equation}
\label{eq:conformal_generators}
\begin{array}{lll}
\mbox{({Translations})} & P_\mu & = J_{5\mu}+J_{6\mu}\\
\mbox{({Conformal transformations})} & K_\mu &= J_{6\mu} - J_{5\mu}\\
\mbox{({Dilations})} & D & = J_{65}\\
\mbox{({Lorentz transformations})} & M_{\mu\nu} & = J_{\mu\nu}.
\end{array}
\qquad\quad \mu= 0,1,2,3.
\end{equation}
Mack has shown that there are 5 classes of positive energy, $P_0\geq0$, unitary representations of $su(2,2)$, all of which possess a lowest weight $\mu=(d;-j_1,-j_2)$. These are characterised by their Poincar\'{e} content, $m=$mass and $s=$spin resp. helicity, as follows \cite{Mack77}:
\begin{subequations}\label{eq:su22Unirreps}
\begin{align}
&\mbox{\textbf{Trivial}}\nonumber\\
&\qquad\mbox{(1) } d=j_1=j_2=0.\\
&\mbox{\textbf{Massive} }m>0\nonumber\\
&\qquad\mbox{(2) }\,j_1 \neq 0,\,j_2 \neq 0,\, d >j_1 +j_2 + 2\qquad s =|j_1 -j_2|... j_1 +j_2\\
&\qquad\mbox{(3) }\,j_1j_2=0,\,d>j_1+j_2+1\qquad s=j_1+j_2\\
&\qquad\mbox{(4) }\,j_1\neq0, j_2\neq0,\,d =j_1+j_2+2 \qquad s = j_1 +j_2.\\
&\mbox{\textbf{Massless}}\nonumber\\
&\qquad\mbox{(5) }\,j_1j_2=0,\, d =j_1+j_2+1 \qquad \mbox{helicity }= j_1-j_2.\label{eq:masslessConditionsMack}
\end{align}
\end{subequations}
%
The last of these, the class of conformally invariant massless representations, are those on which the operator $P^\mu P_\mu$ acts trivially (starting with this fact we provide in Appendix \ref{subsec:MasslessnessAndBosonicOscillatorModel} an algebraic re-derivation of these massless conditions). The class of positive energy representations may be placed within the broader classification of unitary representations of $su(2,2)$, see for example \cite{Todorov1966,Raczka1966} and in particular \cite{Yao1967,Yao1968}. Unitary representations, each of which comprise an infinite sum of finite-dimensional irreducible representations of $\mathcal{S}$, may be classified as either continuous or discrete representations depending on the allowable values of $j_1,j_2$ and $d$. Within each of these classes exist so-called \emph{degenerate} representations in which distinct representations of $\mathcal{S}$ occurs at most once. Let $|k_1,m_1;k_2,m_2;d'\rangle$, $-k_1\leq m_1\leq k_1$, $-k_2\leq m_2\leq k_2$ be a (weight) basis for an irreducible $\mathcal{S}$-module satisfying
\begin{align*}
	&\boldsymbol{M}^2|k_1,m_1;k_2,m_2;d'\rangle=k_1(k_1+1)|k_1,m_1;k_2,m_2;d'\rangle\\
	&\boldsymbol{N}^2|k_1,m_1;k_2,m_2;d'\rangle=k_2(k_2+1)|k_1,m_1;k_2,m_2;d'\rangle\\
	&H_l|k_1,m_1;k_2,m_2;d'\rangle=m_l|k_1,m_1;k_2,m_2;d'\rangle\qquad l=1,2\\
	&H_0|k_1,m_1;k_2,m_2;d'\rangle=d'|k_1,m_1;k_2,m_2;d'\rangle.	
\end{align*}
A fact that will be crucial to the proof of Lemma \ref{lemma:quadratic_identities} is that for degenerate representations every non-zero weight vector $|k_1,m_1;k_2,m_2;d'\rangle$ occurring in an $\mathcal{S}$-submodule occurs in the $su(2,2)$ representation with unit multiplicity \cite{Yao1967}\footnote{For non-degenerate representations an additional (cubic) operator must be constructed in order to form a maximal set of mutually commuting operators and resolve the labelling of states.}. In the analysis of Yau \cite{Yao1967} representations of $\mathcal{S}$ are labeled by $p=k_1+k_2$, $q=k_1-k_2$ and $d'$. For the \emph{lowest} (in terms of the extremal weight) of all these representations we have the correspondence $k_1\rightarrow -j_1$ and $k_2\rightarrow -j_2$ denoted $p_0=-(j_1+j_2)$ and $q_0=-(j_1-j_2)$. Importantly the class of massless representations are degenerate; they are easily shown to satisfy
\begin{align*}
	&p_0\in \{0,\textstyle{\frac12,1,\frac32},2,...\}\\
	&q= \pm p_0\\
	&d'=p+1
\end{align*}
which are the required conditions to belong, in the classification of Yau, to the series of exceptional (most) degenerate discrete representations (see \S 4 of \cite{Yao1968})\footnote{For massless representations the conditions $d'=k_1+k_2+1$ and $k_1-k_2={\sf const}$ reduces the number of state labels from five to just three. This agrees with the characterisation of the most degenerate representations obtained in \cite{Todorov1966}, see also \cite{Castell1968} \cite{Mack&Todorov1969}.}.


\section{Quadratic Superconformal Algebra \texorpdfstring{$su_2(2,2/1)$}{}}\label{sec:QSuperconformalAlgebra}
We begin this section with a review of the general theory of characteristic identities, specialising to $gl(n)$ and its real forms, and finally to a detailed analysis of the minimal identities satisfied by a certain matrix array of $u(2,2)$ generators. Next we define the quadratic superconformal algebra $su_2(2,2/1)$, a real form of $gl_2(4/1)$, where our  aim is to determine which if any representations of the even subalgebra allow the anticommutator $\{\overline{Q},Q\}$ to be brought into correspondence with the minimal identity. In other words we aim to construct an algebraic deformation of the standard superconformal algebra in which $\{\overline{Q},Q\}$ is \emph{identically zero} for a certain subset of representations induced from unitary irreducible representations of the even subalgebra $u(2,2)$. In such a model those representations satisfying this special condition, each of which corresponds to a unique particle in the standard model, would have the remarkable property of carrying supersymmetry without the need for a superpartner. 

\subsection{Review of characteristic identities of Lie algebras}
For $L$ a semi-simple Lie algebra, let $V(\varphi)$ be a fixed finite-dimensional $L$-module with distinct weights $\varphi_i$, $i=1,...,k$. Let $V(\mu)$ be another, possibly infinite-dimensional, $L$-module admitting a character $\chi_\mu:Z\rightarrow\mathbb{C}$, where $Z$ is the centre of $U(L)$. Associated with $V(\varphi)$ and $V(\mu)$ are the representations $\pi_\varphi$ and $\pi_\mu$ respectively. We define
\begin{align*}
	Y\equiv V(\varphi)\otimes V(\mu)
\end{align*}
and for fixed $z\in Z$ we introduce the following object belonging to $($End$\,V(\varphi))\otimes($End$\,V(\mu))$:
\begin{align}\label{eq:matrixArrayA}
	A \equiv A^{\varphi}_{\mu}(z)\equiv-\frac12\left[\partial(z)-\pi_\varphi(z)\otimes1-1\otimes \pi_{\mu}(z)\right]
\end{align}
where $\partial(x)=\pi_{\varphi}(x)\otimes {\sf id} + {\sf id}\otimes \pi_{\mu}(x)$ for $x\in L$ extended to an algebra homomorphism for all of $U$ .

Following Gould \cite{gould1984} (see also \cite{Kostant1975}) the key idea in the determining the minimal polynomial identity satisfied by $A$ is that, under the action of $A$, $Y$ admits a  decomposition into generalised eigenspaces $Y_i$, $i=1,..,m<k$, where $m$ is the number of distinct characters amongst $\chi_{\mu+\varphi_i}$, $i=1,...,k$. The corresponding (generalised) eigenvalues, here called the \emph{characteristic roots}, are 
\begin{equation}\label{eq:A_eigenvalues}
a_i=-\frac{1}{2}\left[\chi_{\mu+\lambda_i}(z)-\chi_{\mu}(z)-\chi_{\lambda}(z)\right]\qquad i=1,..,k.
\end{equation}
Let $m_i$, $i=1,..,m$ be the multiplicity of the $\chi_{\mu+\varphi_i}$ in the original set $\chi_{\mu+\varphi_i}$, $i=1,...,k$. We have
\begin{align*}\label{eq:Y_i_definition}
	Y_i = \{|\eta\rangle\in Y|(A-a_i)^{n_i}|\eta\rangle=0\quad i=1,..,m\}.
\end{align*}
where the integer $n_i\leq m_i $ is the maximal rank of the generalised eigenvectors belonging to $Y_i$. It follows, given $Y =\bigoplus_{i=1}^n Y_i$, that $A$ satisfies the minimal identity
\begin{equation}
m(A)=\prod_{i=1}^m (A- a_i)^{n_i}=0.
\label{eq:minimal_polynomial_identity}
\end{equation}
of degree $\sum{n_i}\leq k$. While $A$ always satisfies an identity of degree $k$, obtained by replacing $n_i$ with $m_i$ in the equation above, it is in practice a difficult problem to obtain the minimal identity since one must determine the maximal rank $n_i$ for each subspace. The situation simplifies greatly when both $V(\varphi)$ and $V(\mu)$ are finite-dimensional irreducible representations. In this case, owing to complete reducibility, each $Y_i$ is itself a finite-dimensional representation which character $\chi_{\mu+\varphi_i}$. Hence $n_i=0$ except when $\mu+\varphi_i$ is dominant integral in which case $n_i=1$.  

Another case which simplifies the calculation of $m(A)$ is when each of the characters $\chi_{\mu+\varphi_i}(z)$ is distinct. In this case $Y_i$, $i=1,...,k$ are again ordinary eigenspaces with $n_i\leq1$ and it remains only to determine which if any $Y_i=(0)$.

Finally we comment on the calculation of the roots \eqref{eq:A_eigenvalues}. We fix $z$ to be the universal Casimir element ${\sf C_2}$. In the case of characteristic modules with a \emph{highest weight} $\lambda'$ we have the textbook rule $\chi_{\lambda'}({\sf C_2}) = \langle\lambda',\lambda'+2\rho\rangle$, where $\rho$ is the half sum of the positive roots of $L$ and $\langle\,,\,\rangle:H^*\times H^*\rightarrow\mathbb{C}$ is the restriction of the Killing form to $H^*$. In practice this is most easily evaluated if the roots are expressed in the Cartesian basis $\varepsilon_i$, $i=1,..,n$ satisfying $\langle\varepsilon_i,\varepsilon_j\rangle=\delta_{ij}$. It will also be necessary for us to calculate the roots for representations characterised by a \emph{lowest weight} $\lambda$. In this case it is straight-forward to derive the corresponding rule\footnote{Take the standard Cartan-Weyl basis $h_\alpha$, $e_\alpha$, $f_\alpha$ with $[e_\alpha, f_\alpha]=h_\alpha$, where $\alpha_i$ and $\alpha$ are the simple and positive roots respectively, and consider the action of ${\sf C_2}=x^ix_i=h^{\alpha_i}h_{\alpha_i}+\sum_{\alpha}(-h_\alpha+2e_\alpha f_\alpha)$ on a lowest weight vector $v_\lambda$.},
\begin{align}\label{eq:lowestWeightRule}
\chi_\lambda({\sf C_2}) = \langle\lambda,\lambda-2\rho\rangle\qquad\mbox{($\lambda$ is the lowest weight).}
\end{align}

\subsection{Minimal identities of \texorpdfstring{$u(2,2)$}{}}

Take $L=gl(n)$ in the Gel'fand basis. We set $z={\sf C_2}=E^i{}_jE^j{}_i$, $V(\varphi)=\{\overline{1}\}$ the fundamental contragredient module with lowest weight $\varphi=-\varepsilon_1$ and $V(\mu)$ an arbitrary lowest weight $L$-module with lowest weight $\mu=\sum_{i=1}^n{\mu_i\varepsilon_i}$. The matrix array \eqref{eq:matrixArrayA}, here denoted by $E$ instead of $A$, takes the form
\begin{align}
	E&=-\pi_{\varphi}(E^j{}_i)\otimes \pi_{\mu}(E^i{}_j)\nonumber\\	
	&=e^i{}_j\otimes \pi_{\mu}(E^i{}_j).
\end{align}
Thus $E$ is simply the standard matrix array of Gel'fand generators in the representation $\pi_\mu$. We abuse notation denoting by $E^{i}{}_j$ both the abstract Gel'fand generator and also the $(i,j)$-matrix element of the array $E$. We recall that the $n$-linearly independent Casimir elements generating the centre of $U(gl(n))$ can be written ${\sf C_n}=\langle E^n\rangle$ where the matrix powers are defined in the obvious way $(E^k)^a{}_b=(E^{k-1})^a{}_cE^c{}_b$ and $\langle\,\,\rangle$ denotes the trace.

Using \eqref{eq:lowestWeightRule} the characteristic roots \eqref{eq:A_eigenvalues} are
\begin{align*}
	a_i&= -\frac12\left(\langle\mu+\varphi_i,\mu+\varphi_i-2\rho\rangle - \langle\mu,\mu-2\rho\rangle -\langle\varphi,\varphi-2\rho\rangle\right)\nonumber\\
	&=\frac12\langle\varphi,\varphi\rangle - \frac12\langle\varphi_i,\varphi_i\rangle + \langle\varphi-\varphi_i,-\rho\rangle - \langle\varphi_i,\mu\rangle.
\end{align*}
The distinct weights of $V(\varphi)$ are $\varphi_i=-\varepsilon_i$ $i=1,..,n$ ordered lowest to highest. It follows that $\langle\varphi,\varphi\rangle = \langle\varphi_i,\varphi_i\rangle$. Writing $\rho=\frac12\sum_{i\geq j}{\varepsilon_i-\varepsilon_j}=\frac12\sum_{i=1}^n{(n+1-2i)\varepsilon_i}$ it is straightforward to calculate
\begin{align}\label{eq:char_roots}
		a_i=\mu_i+i-1.
\end{align}
We saw in \S \ref{sec:superconformalAlgebra} that the basis $E^a{}_b$ $a,b=1,..,4$ for $gl(4)$ may equally be viewed as a basis for the non-compact real form $u(2,2)$ subject, in the case of unitary representations, to the Hermiticity conditions \eqref{eq:hermiticityConditions}. We can therefore, in the present context, view $E$ as an array of $u(2,2)$ generators and use the results above to determine the corresponding characteristic identities. In contrast to $gl(n)$, the determination of the minimal identity is this non-compact case is more complicated due to the non-unitarity of $V(\varphi)$ and the infinite-dimensionality of the positive energy unitary representations $V(\mu)$. 

Using \eqref{eq:cartesianComponentsMu} we may write without any loss of generality the Cartesian components of an arbitrary lowest weight $\mu$ as
\begin{align}\label{eq:weight_cartesian}\begin{array}{rlrl}
	\mu_1 &= k & \mu_3  &= -d+j_1-j_2+k\\
	\mu_2 &= 2j_1 +k \qquad\qquad\qquad&\mu_4 &= -d+j_1+j_2+k,
	\end{array}
\end{align}
where $k$ is related to the eigenvalue of the $gl(1)$ generator \eqref{eq:linear_Casimir_su22}
\begin{align}\label{eq:ZEigenvalue}
	Z=-\langle E\rangle\longrightarrow-\sum\mu_i=2d-4j_1-4k. 
\end{align}
In terms of these labels the characteristic roots \eqref{eq:char_roots} are
\begin{align}\label{eq:roots}
\begin{array}{rlrl}
	a_1 &= k & a_3 &= -d+j_1-j_2+k+2\\
	a_2 &= 2j_1 +k+1 \qquad\qquad\qquad &a_4 &= -d+j_1+j_2+k+3.
\end{array}
\end{align}
We proceed by investigating the existence of a lowest weight vector in each of the subspaces $Y_i$. If $Y_i$ is nonzero then it must, given that $Y$ is the product of two lowest weight modules, contain a lowest weight vector $|\Lambda_i\rangle$ of weight $\mu+\varphi_i$ \cite{Kostant1975}.

	\begin{lemma}[Quadratic polynomial identities of $u(2,2)$]\label{lemma:quadratic_identities}
Take $\mu$, as given by \eqref{eq:weight_cartesian}, to be the lowest weight of a unitary representation of $u(2,2)$ which under the restriction to $su(2,2)$ satisfies the massless conditions \eqref{eq:masslessConditionsMack}. For $\varphi$ the \emph{lowest} weight of the fundamental contragredient representation the matrix array $E=\pi_{\varphi}(E^i{}_k)\otimes\pi_{\mu}(E^k{}_j)$ satisfies the following quadratic identities:
\begin{subequations}\label{eq:u22_quadratic_identities}
\begin{align}
	&\quad(E-a_1)(E-a_3)=0\qquad\qquad\left(\mbox{when }j_1=0\right)\\
	&\quad(E-a_1)(E-a_2)=0\qquad\qquad\left(\mbox{when }j_2=0\right),
\end{align}
\end{subequations}
where
\begin{align*}
	a_i=\mu_i+i-1.
\end{align*}

	\end{lemma}
	\noindent\textbf{Proof.}
	$E$ generically satisfies the quartic identity $\prod_{i=1}^4(E-a_i)$. The proof consists of analysing the decomposition of $Y\equiv V(\varphi)\otimes V(\mu)$ to determine the conditions for which the subspaces $Y_i$ are trivial. For each trivial subspace the degree of the minimal identity satisfied by $E$ is reduced by one where the new identity is obtained from the quartic identity by deletion of the corresponding factors. 
	
For a subspace $Y_i$, $i=1,2,3,4$ to be non-trivial it must contain a lowest weight vector of weight $\mu+\varphi_i$ where $\varphi_i=-\varepsilon_i$ are the weights of $V(\varphi)$ ordered from lowest to highest. Let $|-\varepsilon_i\rangle$ denote the corresponding weight vectors satisfying $E^a{}_b|-\varepsilon_c\rangle=-\delta^a{}_c|-\varepsilon_b\rangle$. For each $Y_i$ we introduce candidate lowest weight vectors $|\Lambda_i\rangle$ consisting of the most general linear combination of all vectors $|\varphi'\rangle\otimes|\mu'\rangle\in Y$ with weight $\mu+\varphi_i$. Since the massless representations of $su(2,2)$ are degenerate each weight vector $|\mu'\rangle$ occurs with unit multiplicity and is uniquely characterised by the eigenvalues $m_1$, $m_2$ and $d'$ of $H_1$, $H_2$ and $H_0$ respectively. This significantly reduces the number of available terms when compared with the general non-degenerate case, viz
	\begin{align*}
	&|\Lambda_1\rangle\equiv |-\varepsilon_1\rangle\otimes|\mu\rangle\\
	&|\Lambda_2\rangle\equiv {\sf a}_1|-\varepsilon_2	\rangle\otimes|\mu\rangle+
			{\sf a}_2(|-\varepsilon_1\rangle\otimes E^1{}_2|\mu\rangle)\\
	&|\Lambda_3\rangle\equiv {\sf b}_1|-\varepsilon_3\rangle\otimes|\mu\rangle
			+{\sf b}_2(|-\varepsilon_2\rangle\otimes E^2{}_3|\mu\rangle)\nonumber+
			{\sf b}_3(|-\varepsilon_1\rangle\otimes E^1{}_3|\mu\rangle)\\
	&|\Lambda_4\rangle\equiv {\sf c}_1 |-\varepsilon_4\rangle\otimes|\mu\rangle	
			+{\sf c}_2(|-\varepsilon_3\rangle\otimes E^3{}_4|\mu\rangle)\nonumber\\
			&\qquad\qquad+{\sf c}_3(|-\varepsilon_2\rangle\otimes E^2{}_4|\mu\rangle)
			+{\sf c}_4(|-\varepsilon_1\rangle\otimes E^1{}_4|\mu\rangle)\nonumber
		\end{align*}
To determine the existence of non-trivial values for the constants above we demand that the simple negative root vectors $E^{a+1}{}_a$, $a=1,2,3$ annihilate each of $|\Lambda_i\rangle$. Clearly $E^{a+1}{}_a|\Lambda_1\rangle=0$, $a=1,2,3$, hence $Y_1$ is always non-trivial. For $|\Lambda_2\rangle$ we have
\begin{align*}
	E^2{}_1|\Lambda_2\rangle&=-{\sf a}_1|-\varepsilon_1	\rangle\otimes|\mu\rangle+
	{\sf a}_2(|-\varepsilon_1\rangle\otimes E^2{}_1E^1{}_2|\mu\rangle)\\
&=-({\sf a}_1+2{\sf a}_2j_1)|-\varepsilon_1	\rangle\otimes|\mu\rangle\\
E^3{}_2|\Lambda_2\rangle&=E^4{}_3|\Lambda_2\rangle=0.
\end{align*}
It follows immediately that $Y_2$ is non-trivial if and only if $j_1\neq0$ which implies $j_2=0$. Proceeding similarly for $|\Lambda_3\rangle$,
\begin{align*}
	E^2{}_1|\Lambda_3\rangle&=-{\sf b}_2|-\varepsilon_1	\rangle\otimes E^2{}_3|\mu\rangle+
	{\sf b}_3(|-\varepsilon_1\rangle\otimes E^2{}_1E^1{}_3|\mu\rangle)\\
&=-({\sf b}_2-{\sf b}_3)|-\varepsilon_1	\rangle\otimes E^2{}_3|\mu\rangle\\
E^3{}_2|\Lambda_3\rangle&=-{\sf b}_1|-\varepsilon_2\rangle\otimes|\mu\rangle
			+{\sf b}_2(|-\varepsilon_2\rangle\otimes E^3{}_2E^2{}_3|\mu\rangle)\nonumber+
			{\sf b}_3(|-\varepsilon_1\rangle\otimes E^3{}_2E^1{}_3|\mu\rangle)\\
			&=-({\sf b}_1+{\sf b}_2(d-j_1+j_2))|-\varepsilon_2\rangle\otimes|\mu\rangle-
			{\sf b}_3(|-\varepsilon_1\rangle\otimes E^1{}_2|\mu\rangle)\\
E^4{}_3|\Lambda_3\rangle&=0,			
\end{align*}
where for non-trivial $Y_3$, $E^3{}_2|\Lambda_3\rangle=0$ if and only if $E^1{}_2|\mu\rangle=0$ which implies $j_1=0$. Finally it is straight-forward to show that $Y_4$ is always trivial, in particular
\begin{align*}
	E^4{}_3|\Lambda_4\rangle&=-{\sf c}_1 |-\varepsilon_3\rangle\otimes|\mu\rangle	
			+{\sf c}_2(|-\varepsilon_3\rangle\otimes E^4{}_3E^3{}_4|\mu\rangle)\nonumber\\
			&\qquad\qquad+{\sf c}_3(|-\varepsilon_2\rangle\otimes E^4{}_3E^2{}_4|\mu\rangle)
			+{\sf c}_4(|-\varepsilon_1\rangle\otimes E^4{}_3E^1{}_4|\mu\rangle)\\
			&\neq0.
\end{align*}
Using the above results we have for $j_2=0$ and $j_1=0$ the decompositions $Y=Y_1\oplus Y_2$ and $Y=Y_1\oplus Y_3$ respectively.\qed\\


We introduce now the quadratic superalgebra $su_2(2,2/1)^{\alpha,\,{\sf c}}$, a real form of $gl_2(4/1)^{\alpha,\,{\sf c}}$, satisfying the commutation relations \eqref{eq:comm_relns_su22_a} and \eqref{eq:comm_relns_su22_b} together with \eqref{eq:generic_solution}. In light of Lemma \ref{lemma:contraction_limit} we may view $su_2(2,2/1)^{\alpha,\,{\sf c}}$ for ${\sf c}=0$ as a quadratic deformation of the ordinary superconformal algebra $su(2,2/1)$. We investigate the possiblity of expressing the quadratic right-hand-side of the anticommutator in terms of the minimal identities \eqref{eq:u22_quadratic_identities}. We continue in the Gel'fand basis satifying \eqref{eq:gelfand_relations} and \eqref{eq:adActionOdd}.
	
	\begin{lemma}
	The anticommutator for $su_2(2,2/1)^{\alpha,\,{\sf c}}$ may be brought into correspondence with the quadratic identities \eqref{eq:u22_quadratic_identities} for massless representations of the even subalgebra. The coincidence occurs for the parameter value $\alpha=-3$ and for central charge ${\sf c}=0$.
	\end{lemma}
	
	\noindent\textbf{Proof.} We begin by noting that for $\alpha=-3$ the anticommutator \eqref{eq:generic_solution} may be expressed in the form
	\begin{align*}
		 \{\overline{Q}^a,Q_b\}=p(E)^a{}_b+\delta^a{}_b {\sf Tr}[p(E)].
	\end{align*}
It remains to show that $p(E)=E(E-(\langle E\rangle +3))$ coincides with the quadratic identities \eqref{eq:u22_quadratic_identities}. Let us start with the $j_1=0$ case; we require
\[(E-a_1)(E-a_3)=	E(E-(\langle E\rangle +3))\]
where from \eqref{eq:roots} and \eqref{eq:ZEigenvalue} we have
\begin{align*}
	a_1=k,\quad a_3=-d-j_2+k+2\quad
\mbox{and}\quad\langle E\rangle=-2d.
\end{align*}
Setting $a_1=k=0$ it follows immediately that $a_3=\langle E\rangle +3$ for $d=j_2+1$. Similarly for the	 $j_2=0$ case we require
\[(E-a_1)(E-a_2)=	E(E-(\langle E\rangle +3))\]
where from \eqref{eq:roots} and \eqref{eq:ZEigenvalue} we have
\begin{align*}
	a_1=k,\quad a_2=2j_1+k+1\quad
\mbox{and}\quad\langle E\rangle=-2d+4j_1+4k.
\end{align*}
Setting $a_1=k=0$ it follows immediately that $a_2=\langle E\rangle +3$ for $d=j_1+1$.	
\qed\\


\section{Conclusion}
\label{sec:Conclusion}
In this paper we have provided a detailed investigation of the structure and properties of a specific member of the class of
quadratic superalgebras introduced by us in J Phys A 44(23):235205 (2011), \cite{JarvisRudolphYates2011}, as a candidate generalized space-time symmetry: namely, the quadratic superalgebra $su_2(2,2/1)$\, a real form of the general class 
$gl_2(n/1)$ for case $n=4$, which can be seen as an extension of the conformal Lie superalgebra $su(2,2/1)$.

In particular, we consider unitary irreducible representations of the conformal algebra $su(2,2)$ (the Lie algebra of the conformal group $SU(2,2)\cong SO(4,2)$ acting on Minkowski space in four dimensions). We have shown that,
for each positive energy, massless unitary irreducible representation, the $su(2,2)$ Lie algebra generators possess a quadratic minimal characteristic identity, which coincides with the anticommutator bracket of the odd generators of the quadratic conformal superalgebra, for a specific choice of free parameter and central charge, which identifies a unique member of the class.  Thus all positive energy, massless unitary irreducible representations of the conformal algebra (which descend to massless unitary irreducible representations of the Poincar\'{e} subgroup, describing massless particle fields), are at the same time, `zero-step'  irreducible representations of the quadratic superconformal algebra: the supersymmetry generators are identically zero, thus obviating the need for super-partners. Thus, if the supersymmetry generators indeed correspond to conserved charges, then under this mechanism, supersymmetry is \emph{unbroken} in an extreme way: the generators annihilate \emph{all} physical states, not only the vacuum state (as in the usual analysis of symmetry breaking \cite{CooperFreedman1983,Witten1981}).

In this investigation we have been at pains to establish the appropriate foundations for our results,
firstly by independently verifying the important PBW property for the enveloping algebra, and then, through a detailed examination of the characteristic identity derivation, by establishing the validity of the quadratic minimal polynomial of $u(2,2)$ for the massless case. As pointed out already in \cite{JarvisRudolphYates2011}, the structure of the quadratic superalgebra
resembles instances of $W$-superalgebras (see for example \cite{deBoer1993, Ragoucy1999}); however our analysis is independent of any reference to the related formalisms and would also apply, for instance, to quadratic extensions of
Poincar\'{e} supersymmetry, or even to some Galilean or nonrelativistic space-time symmetry algebras.
The PBW property alone however is not sufficient to guarantee the existence of any valid coproduct, thus the development of a true `representation theory' must be left for future work. On the other hand, $su_2(2,2/1)$ has $su(2,2/1)$ as a contraction limit, and to this extent, as a generalized supersymmetry its relation to standard supersymmetry is analogous to that of conformal symmetry to Poincar\'{e} symmetry, for example. 

Our results while algebraically framed, currently lack a `microscopic' or model foundation. As indicated in the introduction, and evident from \cite{HLS1975} for example, any non-standard supersymmetry of the type we propose must invoke some extension of the postulates of relativistic quantum fields. In this context it is useful to pursue the analogy with colour symmetry in the spectrum of strongly interacting particles: all particles have colour zero (that is, the colour charge annihilates all physical states), so that colour is a `hidden symmetry'. In terms of free field quark current algebras, extending colour singlet states from mesons to trilocal colour singlet baryons (see \cite{DelbourgoSalam1965}) clearly entrains a type of quadratic superalgebra. Similar considerations arise in the classification of colour singlet operators in a Hamiltonian lattice approach to QCD
\cite{Jarvis&Rudolph2003,RudolphJarvisKijowski2005}. \\[1cm]

\noindent
\textbf{Acknowledgements:}\\
PDJ acknowledges travel support from the Alexander von Humboldt Foundation. We thank Prof G Rudolph for discussions and comments relating to the early development of this work. Part of the work of LAY has been undertaken under the award of an Australian Commonwealth APA postgraduate scholarship. LAY would also like to acknowledge the support of Elizabeth College, Department of Education, Tasmania.\\

\vfill
\pagebreak

\begin{appendix}
\setcounter{equation}{0}
\renewcommand{\theequation}{{A}-\arabic{equation}}
\section{Appendix}
\label{sec:appendix}
\subsection{\texorpdfstring{$gl_2(n/1)^{\alpha,{\sf c}}$}{} structure constants and isomorphisms}
\label{subsec:Isomorphisms}
For completeness we here reiterate from \cite{Jarvis&Rudolph2003,JarvisRudolphYates2011} the parametric form of the quadratic superalgebra $gl_2(n/1)$, and illustrate various scaling and other constraints determining isomorphic forms. Recall that the most general nontrivial odd anticommutator bracket compatible with $gl(n)$ covariance is given by the expansion
\begin{align}\label{eq:anticomm_gln_covariant}
{\{}\overline{Q}^i, Q_j {\}} = & \, 
\texttt{a} (E^2)^i{}_j + E^i{}_j\big( \texttt{b}_{1}  \langle E \rangle +   \texttt{b}_{2}   \big) 
+ \delta^i{}_j \big( \texttt{c}_{1}\langle E^2 \rangle + \texttt{c}_{2}\langle E \rangle^2 + \texttt{c}_{3}\langle E \rangle \big)
+ \texttt{c}\delta^i{}_j \,,
\end{align} 
and that the Jacobi identity imposes the 1-parameter form given in (\ref{eq:generic_solution}) above.
The following is verified by explicit computation:\\[.5cm]

\begin{lemma}[\textbf{Shifting isomorphism}]\mbox{}\\\label{lemma:ShiftingIsomorphism}
Let $gl_2(n/1)^{\alpha,\texttt{c}}$ be the quadratic superalgebra with bracket relations (\ref{eq:generic_solution})
\begin{align*}
{\{}\overline{Q}{}^a, Q_b{\}} = 
(E^2){}^a{}_b - E{}^a{}_b\big(\langle E \rangle - \alpha\big)
-&\,\textstyle{\frac 12}\delta{}^a{}_b\big(\langle E^2 \rangle \!-\! \langle E \rangle^2 \!+\! (n\!-\!1\!+2\alpha)\langle E \rangle \big) 
+ {\texttt c}\delta^a{}_b\,.
\end{align*}
(where an overall rescaling to  $\texttt{a}=1$ is understood). Define 
\[
\sigma\big(E^a{}_b\big) = E^a{}_b - \frac{\lambda}{n}\delta^a{}_b \mathbb{I}\,, \qquad \sigma\big(\overline{Q}^a\big) = \overline{Q}^a\,,\qquad
\sigma \big({Q}_b\big)=  {Q}_b\,,
\]
Then we have an isomorphism of algebras $gl_2(n/1)^{\alpha,\texttt{c}} \cong gl_2(n/1)^{\sigma(\alpha),\sigma(\texttt{c})}$
with
\[
\sigma(\alpha) = \alpha - (n-2)\frac{\lambda}{n}\,,\qquad \sigma(\texttt{c}) = c -(n-1)\frac{\lambda}{n}
(\alpha - \textstyle{{\frac 12}}(n-2)\displaystyle{\frac{\lambda}{n}}+\textstyle{{\frac 12}}n)  \,.
\]
In particular for the choice $\lambda/n =\alpha/(n-2)$, we have  
\[
gl_2(n/1)^{\alpha,\texttt{c}} \cong gl_2(n/1)^{0,\overline{\texttt{c}}}\,,
\qquad \overline{\texttt{c}}= \texttt{c}-\textstyle{{\frac 12}}\displaystyle{\frac{(n-1)}{(n-2)}}\alpha(\alpha+ n)\,.
\]
\hfill $\Box$
\end{lemma}

\noindent
A slightly more general form arises if the odd generators transform as \emph{densities} under $gl(n)$, rather than vector operators as we have assumed hirtherto. Re-evaluating the Jacobi constraints on the above structure coefficients  
$\texttt{b}_1, \cdots, \texttt{c}_3$ leads to: \\

\begin{lemma}[\textbf{Weighted quadratic superalgebra}]\mbox{}\\\label{lemma:WeightedQSA}
Under the modified $gl(n)$ transformation properties with the odd generators being densities of weight $w$\,,
\[
{[}E^a{}_b, \overline{Q}^c{]} = \delta_b{}^c \overline{Q}^a - w \delta_b{}^a \overline{Q}^c\,, \qquad 
{[}E^a{}_b, {Q}_c{]} = -\delta^a{}_c {Q}_b+ w\delta^a{}_b \overline{Q}^c\,,
\]
the Jacobi constraints on the $gl_2(n/1)$ structure coefficients (as in eq. (\ref{eq:generic_solution}), Lemma \ref{lemma:ShiftingIsomorphism} above)
$\texttt{b}_1, \cdots, \texttt{c}_3$ lead to the following parametric solution
\begin{align}\label{eq:weighted_anticommutator}
{\{}\overline{Q}{}^a, Q_b{\}} = &\,
(E^2){}^a{}_b \!-\! E{}^a{}_b\left(\displaystyle{\frac{1\!-\!2w}{1\!-\!nw}}\langle E \rangle \!-\! \alpha\right) \\
&\,-\textstyle{\frac 12}\delta{}^a{}_b\left(\langle E^2 \rangle -
\displaystyle{\frac{1\!-\!4w\!+\!(n\!+\!2)w^2}{(1\!-\!nw)^2}}\langle E \rangle^2 +
\big(\displaystyle{\frac{n\!-\!1\!+\!5w\!-\!3nw^2}{1\!-\!nw}}\!+\!2\frac{1\!-\!w}{1\!-\!nw}\alpha \big)\langle E \rangle \right)
+ {\texttt c}\delta^a{}_b\,.\nonumber
\end{align}
\mbox{}\hfill $\Box$
\end{lemma}

\noindent
A further possible quadratic $gl_2(n/1)$ superalgebra structure arises in the presence of 
additional odd-graded quadratic constraints, of the general form

\begin{align}\label{eq:aux_relations}
	S_k E^k{}_i =  S_i \big({\sf s} \langle E \rangle + {\sf t}\big)\,,\qquad 
	E^i{}_k\overline{S}^k = \big({\sf s} \langle E \rangle + {\sf t}\big)\overline{S}^i\,.
\end{align}
The effect of these identities is to reduce the number of distinct index types arising in the Jacobi identity constraints (in that previous separate conditions become combined through linear combinations with the fixed parameters 
${\sf s}$, ${\sf t}$\,). We have

\begin{lemma}[\textbf{Flexible Quadratic Superalgebra:}]\mbox{}\\\label{lemma:FlexibleQSA}
The anticommutator bracket structure with weighted generators is relaxed with respect to the Jacobi identities in the presence of additional odd-graded quadratic constraints as above (see (\ref{eq:aux_relations}). While coefficient $\texttt{b}_1$ is unchanged
from (\ref{eq:weighted_anticommutator}), namely 
\[
\texttt{b}_1 = -\displaystyle{\frac{1\!-\!2w}{1\!-\!nw}}\,,
\]
there remain only 2 constraints on the 4 remaining coefficients $\texttt{c}_1,  \texttt{c}_2, \texttt{c}_3$
(with $\texttt{b}_2:= \alpha$ undertermined)\,
\begin{align*}
(1-w)\texttt{b}_1 +  {\sf s}  =&\, (2w-2{\sf s} ) \texttt{c}_1  -2(1-nw)\texttt{c}_2 \,. \\
(1-w)\texttt{b}_2 + ( {\sf t}+w^2) = &\,(-2{\sf t}+ n-w(2nw-1))\texttt{c}_1 + (1+nw(nw-2))\texttt{c}_{2}
+ (nw-1)\texttt{c}_{3}\,,
\end{align*}
leading to a 2-parameter solution. \hfill $\Box$
\end{lemma}

\noindent
\textbf{Fermionic oscillator realization}\\
As a concrete case study of a quadratic $gl_2(n/1)$ superalgebra (with weighted generators as above) we present %
a construction for the $n=4$ case using standard fermionic oscillators (equivalent to an 8-dimensional Clifford algebra)
$a_i\,, a^j:= (a_j)^\dagger$\,,  with anticommutation relations
\[
\{a_i, a^j \} = \delta_i{}^j\,, \qquad \{a_i, a_j \} =\{a^i, a^j \} =0\,,\qquad  i,j=1,\cdots,4\,.
\]
We define the $gl(n)$ generators
\[
E^i{}_j = a^i a_j\,,
\]
and odd generators with the use of the totally antisymmetric Levi-Civita tensor,
\begin{align*}
\overline{S}^i =&\, \textstyle{\frac 16}\varepsilon^{ik\ell m}a_{k}a_\ell a_m\,,\\
S_j =&\, \textstyle{\frac 16}\varepsilon_{jpqr} a^{k}a^{\ell}a^{m}\,, \qquad k,\ell, m =1,\cdots,4\,.
\end{align*}
for which the commutation and anticommutation relations are
\begin{align*}
{[} E^i{}_j, E^k{}_\ell {]}= &\, \delta_j{}^k E^i{}_\ell - \delta^i{}_\ell E^k{}_j \,, \\
 {[}E^i{}_j, \overline{S}^k{]} = &\, \delta_j{}^k \overline{S}^i -  \delta_j{}^i \overline{S}^k\,, \\
{[}E^i{}_j, {S}_k{]} = &\, -\delta^i{}_k {S}_j+ \delta^i{}_j \overline{S}^k\,,\\
\mbox{and}\qquad \{\overline{S}^i, S_j \}=&\, E^i{}_j(\langle E\rangle -2) +\delta^i{}_j(-\textstyle{\frac 12}\langle E\rangle^2+\textstyle{\frac 32}\langle E\rangle)-\delta^i{}_j\,, \\
\mbox{with}  \qquad \{S_i, S_j \}=&\, \{\overline{S}^i, \overline{S}^j \} =0\,.
\end{align*}

\noindent
Given that the construction is framed in the associative algebra of annihilation and creation modes, the Jacobi identities in this case are of course guaranteed.  Now the above anticommutator bracket
can be written in the general weighted quadratic superalgebra form (Lemma \ref{lemma:WeightedQSA}) in different ways, in consequence of the quadratic characteristic relations
\[
(E^2)^i{}_j =  -E^i{}_j \langle E \rangle + 4 E^i{}_j \,, \qquad \langle E^2 \rangle = - \langle E \rangle^2 + 4 \langle E \rangle
\]
satisfied by the generators in this case, by adding a linear combination of these identities and rescaling. In particular it can be checked that
\[
\{\overline{S}^i, S_j \}= -\textstyle{\frac34}(E^2)^i{}_j +\textstyle{\frac 14}E^i{}_j(\langle E\rangle +4) 
+\delta^i{}_j(\textstyle{\frac 18}\langle E^2\rangle-\textstyle{\frac 38}\langle E\rangle^2 +\langle E\rangle)-\delta^i{}_j\,
\]
is one such equivalent 	. 
Further, the anticommutator bracket structure with weighted generators (here $w=+1$\,) is 
relaxed with respect to the Jacobi identities in the presence of 
additional odd-graded quadratic constraints, of the form given in Lemma \ref{lemma:FlexibleQSA} above,
with ${\sf s}=4$\,, ${\sf t}=0$ in the present realization.

In the present case the Jacobi identity conditions are, after an overall rescaling of each parameter by $-\frac34$ (in addition to $\texttt{b}_{1} = \textstyle{\frac 14}\,$):
\[
 \texttt{c}_1 - \texttt{c}_2=\textstyle{\frac 12};
\qquad -\texttt{c}_1 +3 \texttt{c}_2 +\texttt{c}_3=-\textstyle{\frac 14}\,.
\]
These equations are indeed obeyed by the coefficients 
\[
\texttt{c}_{1} = \textstyle{\frac 18}\,; \qquad
\texttt{c}_{2} = -\textstyle{\frac 38}\,; \qquad
\texttt{c}_{3} = 1\,; \qquad
\]
appearing in the above equivalent form (with $\texttt{b}_{2}\equiv \alpha =-\textstyle{\frac 43} $\,).
\vfill
\pagebreak

\subsection{Quadratic algebras and PBW conditions}
\label{subsec:PBWconditions}
In this appendix we provide the detailed proof of Lemma \ref{lemma:QSLAisPBW}, \S \ref{sec:QuadraticSuperalgebras} above, to establish the PBW property of the $gl_2(n/1)$ enveloping algebra, and for completeness we also include a further discussion on the admissibility of additional covariant quadratic auxiliary constraints that arise naturally in some realizations, such as the case of the fermionic oscillator construction dealt with above in \S \ref{subsec:Isomorphisms}.\\

\noindent\textbf{Proof of Lemma \ref{lemma:QSLAisPBW}} The homogeneous relations, obtained by truncation of \eqref{eq:qlsa:quadraticRelations}, are
\begin{align}\label{eq:QSLAHomogeneousRelations}
	I_2&=\{x_ix_j-x_jx_i,\quad x_iy_p-y_px_i,\quad y_py_q+y_qy_p-d_{pq}^{rs}x_rx_s\}.
\end{align} 
Of these relations the only ones which differ from the ordinary Lie superalgebra case are
\begin{align*}
			y_py_q+y_qy_p-d_{pq}^{rs}x_rx_s.
\end{align*}
The index condition $k$, $l$ precedes $p$, $q$ guarantees that the leading monomial of these relations will always be one of $y_py_q$ or $y_qy_p$ from which it follows that
\begin{align}\label{eq:S2_rigid}
	S_2=\{(i,j)|i\leq j\}\cup\{(i,p)\}\cup\{(p,q)|p<q\}.
\end{align}
Employing Lemma \ref{lemma:diamond} we need only consider the following elements of $T_3$:
 $y_ry_qy_p$, $y_ry_qx_i$, $y_rx_jx_i$ and $x_kx_jx_i$ for $p<q<r$ and $i<j<k$. 
Containing at most one odd generator, the cases $y_rx_jx_i$ and $x_kx_jx_i$ $($schematically $\overline{100}$ and $\overline{000})$ satisfy \eqref{eq:diamondLemma} as a direct result of the commutativity of even-even and even-odd pairs of generators. Employing the map
\begin{align*}
	\pi(y_py_q)=\left\{ \begin{array}{ll}y_py_q\qquad\qquad\qquad&(p<q)\\-y_qy_p+d_{pq}{}^{rs}x_rx_s&(p\geq q)\end{array} \right.
\end{align*}
we examine the remaining two cases in detail. In what follows we evaluate separately the left- and right-hand side of \eqref{eq:diamondLemma}, applying the operators $\pi^{12}$ and $\pi^{23}$ successively until each side is completely ordered and $\pi$ acts trivially. The notation $\cdots{\longrightarrow}$ denotes repeated application of these operators in cases where they act simply to reorder commutative pairs. The notation $\{\,\,\,\}_{(ordered)}$ is used for expressions containing a summation over even elements; it indicates that the generators comprising each term of the sum are ordered appropriately for the corresponding values of the summation indices.   
\begin{align*}
\begin{array}{lll}
\overline{\mathbf{110}}\quad&&\\
	LHS:\,& y_ry_qx_i\,&\stackrel{\pi^{12}}{\rightarrow}\,(-y_qy_r+d_{rq}^{st}x_sx_t)x_i\,\cdots{\longrightarrow}\,
	-x_iy_qy_r + d_{rq}^{st}\{x_sx_tx_i\}_{(ordered)}\\
	RHS:\,& y_ry_qx_i\,&\stackrel{\pi^{23}}{\rightarrow}\,y_rx_iy_q\,\stackrel{\pi^{12}}{\rightarrow}\,x_iy_ry_q
	\,\stackrel{\pi^{23}}{\rightarrow}\,x_i(-y_qy_r+d_{rq}^{st}x_sx_t)\\
	&&\cdots{\longrightarrow}\,-x_iy_qy_r + d_{rq}^{st}\{x_ix_sx_t\}_{(ordered)}\\ &&\\
	\overline{\mathbf{111}}\quad&&\\
	LHS:\,& y_ry_qy_p\,&\stackrel{\pi^{12}}{\rightarrow}\,(-y_qy_r+d_{rq}^{st}x_sx_t)y_p\\
	&&\stackrel{\pi^{23}}{\rightarrow}\,-y_q(-y_py_r+d_{rp}^{st}x_sx_t)+d_{rq}^{st}x_sx_ty_p\\
	&&\,\stackrel{\pi^{12}}{\rightarrow}\,(-y_py_q+d_{qp}^{st}x_sx_t)y_r
	-d_{rp}^{st}x_sy_qx_t+d_{rq}^{st}x_sx_ty_p\\
	&&\stackrel{\pi^{23}}{\rightarrow}\,-y_py_qy_r+d_{qp}^{st}x_sx_ty_r-d_{rp}^{st}x_sx_ty_q+d_{rq}^{st}x_sx_ty_p\\	&& \\
	RHS:\,& y_ry_qy_p\,&\stackrel{\pi^{23}}{\rightarrow}\,y_r(-y_py_q+d_{qp}^{st}x_sx_t)\\
	&&\stackrel{\pi^{12}}{\rightarrow}\,-(-y_py_r+d_{rp}^{st}x_sx_t)y_q + d_{qp}^{st}x_sy_rx_t\\
	&&\stackrel{\pi^{23}}{\rightarrow}\,y_p(-y_qy_r+d_{rq}^{st}x_sx_t) - d_{rp}^{st}x_sx_ty_q + d_{qp}^{st}x_sx_ty_r\\
	&&\stackrel{\pi^{12}}{\rightarrow}\,-y_py_qy_r+d_{rq}^{st}x_sy_px_t - d_{rp}^{st}x_sx_ty_q + d_{qp}^{st}x_sx_ty_r\\
	&&\stackrel{\pi^{23}}{\rightarrow}\,-y_py_qy_r+d_{rq}^{st}x_syx_ty_p - d_{rp}^{st}x_sx_ty_q + d_{qp}^{st}x_sx_ty_r
	\end{array}
\end{align*}
In both cases we have $LHS=RHS$ thus satisfying \eqref{eq:diamondLemma}.\qed\\

Introduced in Lemma \ref{lemma:FlexibleQSA} is a more flexible definition of the quadratic anticommutator where, in the presence of additional relations of the form
\begin{align}\label{eq:extra_relations}
	E^a{}_c \overline{Q}^c=({\sf s}\langle E\rangle+{\sf t})\overline{Q}^a
		\qquad Q_c E^c{}_a = Q_a({\sf s}\langle E\rangle+{\sf t}	),
\end{align}
the ordinary $\mathbb{Z}_2$-graded Jacobi identities are fulfilled by a two-parameter family of algebras. Relations of this form may be satisfied on an representation specific basis such as in the fermionic oscillator realisation given at the end of \S \ref{subsec:Isomorphisms}. It is of course possible impose such relations abstractly as additional defining relations of the quadratic superalgebra, leading in principle to a broader family of quadratic superalgebras under examination. However this move has consequences for the form of generalised Jacobi identities and indeed the existence of a PBW basis.    
\begin{lemma}
The two parameter quadratic algebra arising via the imposition of the relations \eqref{eq:extra_relations} as additional defining relations of $gl_2(n/1)$ is not a PBW algebra.
\end{lemma}
\noindent\textbf{Proof.} 
Following the notation of \S\ref{sec:QuadraticSuperalgebras}, we add to the homogeneous relations \eqref{eq:QSLAHomogeneousRelations} the projection of the extra relations onto $X\otimes X$. We express the resulting homogeneous relations in the Gel'fand basis as
\begin{align}\label{eq:PR_serre}
\begin{array}{ll}
	R=\{&E^i{}_jE^k{}_l-E^k{}_lE^i{}_j,\\
	&E^i{}_j\overline{Q}^k-\overline{Q}^kE^i{}_j,\\
	&E^i{}_jQ_k-Q_kE^i{}_j,\\
	&\overline{Q}^pQ_q+Q_q\overline{Q}^p - (d^p_q)^{kl}_{mn}E^k{}_mE^l{}_n,\\
	&E^a{}_bQ_a-{\sf s}(E^1{}_1+E^2{}_2+\cdots+E^n{}_n)Q_b,\\
	&E^b{}_a\overline{Q}^a-{\sf s}(E^1{}_1+E^2{}_2+\cdots+E^n{}_n)\overline{Q}^b\quad\}.
\end{array}
\end{align}
The leading terms associated with $R$ are:
\begin{align*}
	E^i{}_jE^k{}_l\,\,(i,j)>(k,l),\quad \overline{Q}^pQ_q,\quad E^n{}_n\overline{Q}^b\,\,\mbox{and}\,\,E^n{}_nQ_b.
\end{align*}
In other words the set $S_2$ is the same as the ordinary quadratic Lie superalgebra case (see \eqref{eq:S2_rigid}) except we must remove the $2n$ elements corresponding to $\overline{01}$ leading monomials $E^n{}_n\overline{Q}^b\,\,\mbox{and}\,\,E^n{}_nQ_b$, $b=1,..,n$. To determine the action of $\pi$ we set each expression in \eqref{eq:PR_serre} to zero and rearrange to solve for the leading monomial in each case. (Note: If for a given expression belonging to $R$ one or more of the non-leading terms is the leading term of some other expression then additional substitutions are made until the leading term of each expression is given as a linear sum of non-leading terms.) In particular we have
\begin{align*}
	\pi(E^n{}_n\overline{Q}^b)=\left \{\begin{array}{ll}	 
					\scalebox{0.8}{$\dfrac{1}{\sf s}$}E^b{}_a\overline{Q}^a-(E^1{}_1+E^2{}_2+\cdots+E^{n-1}{}_{n-1})\overline{Q}^b\qquad\qquad&(b\neq n)\vspace{2mm}\\
			\scalebox{0.8}{$\dfrac{1}{{\sf s}-1}$}(E^n{}_1\overline{Q}^1+E^n{}_2\overline{Q}^2+\cdots+E^n{}_{n-1}\overline{Q}^{n-1}) &{}
					\\\qquad\qquad-\scalebox{0.8}{$\dfrac{{\sf s}}{{\sf s}-1}$}(E^1{}_1+E^2{}_2+\cdots+E^{n-1}{}_{n-1})\overline{Q}^n\qquad&(b=n).
	\end{array}\right.
\end{align*}
We need only find a monomial in $T_3$ which fails the condition \eqref{eq:diamondLemma}. We take as candidate monomials those which contain as an adjacent pair one of the quadratic terms $E^n{}_n\overline{Q}^b$ or $E^n{}_nQ_b$. Selecting the monomial $E^n{}_n\overline{Q}^b\overline{Q}^c\in T_3$ (assume $c<b\neq n$), we evaluate the left- and right-hand side of \eqref{eq:diamondLemma}, applying the operators $\pi^{12}$ and $\pi^{23}$ successively until each side is completely ordered and $\pi$ acts trivially,
\begin{align*}\begin{array}{lll}
		LHS:&\,E^n{}_n\overline{Q}^b\overline{Q}^c\,&\stackrel{\pi^{12}}{\rightarrow}\,(\scalebox{0.8}{$\dfrac{1}{\sf s}$}E^b{}_a\overline{Q}^a-(E^1{}_1+E^2{}_2+\cdots+E^{n-1}{}_{n-1})		\overline{Q}^b)\overline{Q}^c\\
	&&\stackrel{\pi^{23}}{\rightarrow}\,\scalebox{0.8}{$\dfrac{1}{\sf s}$}E^b{}_a\overline{Q}^a\overline{Q}^c-(E^1{}_1+E^2{}_2+\cdots+E^{n-1}{}_{n-1})\overline{Q}^c\overline{Q}^b\\
	&&\\
	RHS:&\,E^n{}_n\overline{Q}^b\overline{Q}^c\,&\stackrel{\pi^{23}}{\rightarrow}\,E^n{}_n\overline{Q}^c\overline{Q}^b\\
	&&\stackrel{\pi^{12}}{\rightarrow}\,(\scalebox{0.8}{$\dfrac{1}{\sf s}$}E^c{}_a\overline{Q}^a-(E^1{}_1+E^2{}_2+\cdots+E^{n-1}{}_{n-1})\overline{Q}^c)\overline{Q}^b\\
	&&\,=\,\scalebox{0.8}{$\dfrac{1}{\sf s}$}E^c{}_a\overline{Q}^a\overline{Q}^b-(E^1{}_1+E^2{}_2+\cdots+E^{n-1}{}_{n-1})\overline{Q}^c\overline{Q}^b.
\end{array}\end{align*}
Demanding that $RHS=LHS$ yields the condition,
\begin{align}\label{eq:PBW_condition_serre}
	E^b{}_a\overline{Q}^a\overline{Q}^c=E^c{}_a\overline{Q}^a\overline{Q}^b.
\end{align}
Asides from certain representations specific instances, such as the fermionic oscillator realisation at the end of appendix \ref{subsec:Isomorphisms} or in the case of zero-step modules in which the action of $\overline{Q}^p$ is identically zero, we cannot in the general case expect \eqref{eq:PBW_condition_serre} to be satisfied.\qed \\


\subsection{\texorpdfstring{$su(2,2)$}{} massless conditions and oscillator construction}
\label{subsec:MasslessnessAndBosonicOscillatorModel}
We provide in the first part of this appendix an algebraic re-derivation of the massless conditions \eqref{eq:masslessConditionsMack} starting form the assumption $P^\mu P_\mu=0$ on massless multiplets. In the second part we review various oscillator constructions for $su(2,2)$ and $so(4,2)$. 

Let $\mathcal{M}$ be a massless representation, $u\in U(so(4,2))$ and $|\psi\rangle\in\mathcal{M}$. Since $\mathcal{M}$ is invariant under the action of $so(4,2)$ it follows that 
\begin{align}\label{eq:PSquaredInvariance}
	[u,P^\mu P_\mu]|\psi\rangle=0.
\end{align}
Following \cite{Bracken&Jessup:1982a} we introduce the $so(4,2)$ tensor operators
\begin{align}\label{eq:WTensors}
	W_{AB}=J_{AC}J^C{}_B + J_{BC}J^C{}_A+\frac13g_{AB}J_{CD}J^{CD}
\end{align}
which generate an $20$-dimensional invariant subalgebra $\mathcal{W}\subset U$. Using \eqref{eq:conformal_generators} and \eqref{eq:WTensors} we have
\begin{align*}
	P^\mu P_\mu &= g^{\mu\nu}(J_{5\mu}+J_{6\mu})(J_{5\nu}+J_{6\nu})\nonumber\\
	&=-\frac12W_{55}-\frac12W_{66}-W_{56}
\end{align*}
which together with \eqref{eq:PSquaredInvariance} leads to the following result.
\begin{theorem}[Massless Conditions (\cite{Bracken&Jessup:1982a}, Theorem 3.1)] 
 A module $\mathcal{M}$ is a conformally invariant massless representation if and only if
\begin{align}\label{eq:masslessConditionsB&J}
	W_{AB}|	\psi\rangle=0,\qquad A,B=0,1,2,3,5,6.
\end{align}
\end{theorem} 
We now prove the equivalence of the massless conditions \eqref{eq:masslessConditionsMack} and \eqref{eq:masslessConditionsB&J}.  
The set of 20 linearly independent equations \eqref{eq:masslessConditionsB&J} may be rewritten, taking appropriate linear combinations, in a more convenient form expressed in terms of the physical generators \eqref{eq:conformal_generators}. This new set naturally includes 
\begin{align}\label{eq:PSquared=0}
	P^\mu P_\mu|\psi\rangle=0
\end{align}
and we shall require one other, namely (Eq 3.26 \cite{Bracken&Jessup:1982a})
\begin{align}\label{eq:B&JEq3.26}
	K_\mu P^\mu |\psi\rangle = \left(M_{\mu\nu}M^{\mu\nu}-4iD+4D^2\right)|\psi\rangle.
\end{align}
Let $|\psi_\mu\rangle$ be the lowest weight vector of $\mathcal{M}$. We can establish a constraint condition by demanding that the zero-weight contributions of \eqref{eq:PSquared=0} and \eqref{eq:B&JEq3.26} hold in the case $|\psi\rangle=|\psi_\mu\rangle$.

Using \eqref{eq:conformal_generators}\eqref{eq:so42ToPref} we obtain
\begin{align}
\begin{array}{ll}
	P_0=H_0+(X^+_0-X^-_0)\qquad &K_0=H_0-(X^+_0-X^-_0)\\
	\boldsymbol{P}=(\boldsymbol{M}-\boldsymbol{N})+(\boldsymbol{X}^+-\boldsymbol{X}^-)\qquad 
	& \boldsymbol{K}=(\boldsymbol{M}-\boldsymbol{N})-(\boldsymbol{X}^+-\boldsymbol{X}^-)\\
	D=-i(X^+_0-X^-_0) & D^2=-(X^+_0-X^-_0)^2.
\end{array}
\end{align}
$\boldsymbol{M}^2$ and $\boldsymbol{N}^2$ are the quadratic Casimir for each of the compact $SU(2)$ subgroups respectively. Their eigenvalues are 
\begin{align*}
	\boldsymbol{M}^2 |\psi\rangle = j_1(j_1+1)|\psi\rangle\qquad\boldsymbol{N}^2 |\psi\rangle = j_2(j_2+1)|\psi\rangle.
\end{align*}
Using the commutation relations \eqref{eq:comm_relations_preferred} we evaluate the action
\begin{align*}
	\left[(P_0)^2\right]_0|\psi_\mu\rangle &= \left((H_0)^2 - X^+_0X^-_0 - X^-_0X^+_0\right)|\psi_\mu\rangle\\
	&= \left((H_0)^2 + \textstyle{\frac12} H_0\right)|\psi_\mu\rangle\\
	&= d(d+\textstyle{\frac12})|\psi_\mu\rangle
	\end{align*}
and similarly
\begin{align*}
	\left[\boldsymbol{P}^2\right]_0 |\psi_\mu\rangle &= \left(\boldsymbol{M}^2 - 2\boldsymbol{M}\!\cdot\!\boldsymbol{N}+\boldsymbol{N}^2 +\textstyle{\frac32} H_0\right)|\psi_\mu\rangle\\
	&= (j_1(j_1+1) + j_2(j_2+1) - 2j_1j_2-\textstyle{\frac32}d)|\psi_\mu\rangle
\end{align*}
where $\left[\phantom{x^2}\right]_0$ denotes the zero-weight terms of the relevant expression. The requirement
\[\left[P^\mu P_\mu\right]_0 |\psi_\mu\rangle= \left[(P_0)^2 - \boldsymbol{P}^2\right]_0|\psi_\mu\rangle=0\] 
leads to the condition
\begin{align*}
	d(d-1) - j_1(j_1+1) - j_2(j_2+1) + 2j_1j_2=0.
\end{align*}
Proceeding similarly,
\begin{align*}
	\left[M_{\mu\nu}M^{\mu\nu}\right]_0|\psi_\mu\rangle&=\left[2\left(-(M_{10})^2 -(M_{20})^2 -(M_{30})^2 + (M_{12})^2 + (M_{23})^2 + (M_{31})^2\right)\right]_0|\psi_\mu\rangle\\
	&=\left(-3H_0+2\boldsymbol{M}^2 + 4\boldsymbol{M}\boldsymbol{N} + 2\boldsymbol{N}^2\right)|\psi_\mu\rangle\\
	&=\left(2j_1(j_1+1) + 2j_2(j_2+1) + 4j_1j_2 - 3d\right)|\psi_\mu\rangle\\
	{}&{}\\
	\left[K_0P_0\right]_0 |\psi_\mu\rangle &=\left((H_0)^2 - \textstyle{\frac12} H_0\right)|\psi_\mu\rangle\\
	\left[\boldsymbol{K}.\boldsymbol{P}\right]_0|\psi_\mu\rangle &= \left(-\boldsymbol{M}^2 + 2\boldsymbol{M}\boldsymbol{N}-\boldsymbol{N}^2+\textstyle{\frac32} H_0\right)|\psi_\mu\rangle\\
	\Rightarrow \left[K_\mu P^\mu\right]_0|\psi_\mu\rangle&= \left(d(d-2) + j_1(j_1+1) + j_2(j_2+1) - 2j_1j_2\right)|\psi_\mu\rangle.
\end{align*}
and
\begin{align*}
	\left[D^2\right]_0|\psi_\mu\rangle = -\textstyle{\frac12}H_0|\psi_\mu\rangle = -\textstyle{\frac12} d|\psi_\mu\rangle.
\end{align*}
Employing all of the results above and subtracting \eqref{eq:PSquared=0} from \eqref{eq:B&JEq3.26}, we have 
\begin{align*}
	&\quad\,\left[K^\mu P_\mu - P^\mu P_\mu\right]_0|\psi_\mu\rangle = \left[M_{\mu\nu}M^{\mu\nu}- 4iD + 4D^2\right]_0|\psi_\mu\rangle\\
	&\Rightarrow 2j_1(j_1+1) + 2j_2(j_2+1) - 4j_1j_2 - d = 2j_1(j_1+1) + 2j_2(j_2+1) + 4j_1j_2 - d\\
	&\Rightarrow j_1j_2 =0.
\end{align*}
Substituting $j_1j_2 =0$ into \eqref{eq:PSquared=0} gives two possible cases:
\begin{align*}
	(j_2=0)\qquad &d(d-1) = j_1(j_1+1)\quad\Rightarrow\quad d=j_1+1&\quad\mbox{or}\\
	(j_1=0)\qquad &d(d-1) = j_2(j_2+1)\quad\Rightarrow\quad d=j_2+1,&
\end{align*}
which together imply
\[d=j_1+j_2+1\qquad\mbox{where}\qquad j_1j_2 =0.\]
Thus \eqref{eq:masslessConditionsMack} are necessary conditions for \eqref{eq:masslessConditionsB&J}. We now show that they are sufficient. 

Let $W^*$ be the highest weight of the $\mathcal{W}$. The conditions \eqref{eq:masslessConditionsB&J} will hold if 
\begin{align}\label{eq:W*=0}
	W^*|\psi_\mu\rangle=0.
\end{align}
For on any state $|\psi\rangle=u_+|\psi_\mu\rangle$, $u_+\in U_+$, we have
\[W^*|\psi\rangle=W^*u_+|\psi_\mu\rangle=\left(u_+W^*+[W^*,u^+]\right)|\psi_\mu\rangle=0\]
since $[W^*,u_+]=0$ $\forall\ u_+\in U_+$.
There exists, due to irreducibility, $u_{AB}\in U_-$ such that  $W_{AB}=[u_{AB},W^*]$, and it follows that
\begin{align*}
	W_{AB}|\psi\rangle &=W_{AB}(u_+|\psi_\mu\rangle)\\
	&=u_{AB}(W^*u_+|\psi_\mu\rangle)-W^*(u_{AB}u_+|\psi_\mu\rangle)\\
	&=0-W^*(u_+u_{AB}+[u_{AB},u_+])|\psi_\mu\rangle\\
	&=0
\end{align*}
since $W^*[u_{AB},u_+]|\psi_\mu\rangle=0$ for either $[u_{AB},u_+]\in U_+\oplus U_0$ or $[u_{AB},u_+]\in U_-$.

$\mathcal{W}\backsimeq\boldsymbol{20}$ transforms as an $so(4,2)$ tensor representation of symmetry type $\{\overline{1^2};1^2\}$ with corresponding highest weight $\varepsilon_1+\varepsilon_2-\varepsilon_3-\varepsilon_4$. Identifying $\boldsymbol{15}$ as the adjoint representation we have
\[\boldsymbol{15\cdot15}\backsimeq \boldsymbol{84}+\boldsymbol{20}+\boldsymbol{15}+\boldsymbol{1}.\]
Accordingly we may write a basis for $\mathcal{W}$ that is quadratic in the Gel'fand generators taking the form
\begin{align*}
	W^{ij}_{\,kl}=E^i{}_kE^j{}_l-E^j{}_kE^i{}_l-E^i{}_lE^j{}_k+E^j{}_lE^i{}_k + (diagonal\,terms) - (traces)
\end{align*}
In this basis, we have (up to an overall normalisation) 
\begin{align*}
	W^*=W^{12}_{\,34}=E^1{}_3E^2{}_4-E^2{}_3E^1{}_4.
\end{align*}
Finally we demand,
\begin{align*}
	0&=\langle\psi_\mu|(W^*)^\dagger W^*|\psi_\mu\rangle\\
	&=\langle\psi_\mu|(E^4{}_2E^3{}_1-E^4{}_1E^3{}_2)(E^1{}_3E^2{}_4-E^2{}_3E^1{}_4)|\psi_\mu\rangle\\
	&=\langle\psi_\mu|(E^3{}_3-E^1{}_1)(E^4{}_4-E^2{}_2)+2(E^3{}_3-E^2{}_2)+(E^3{}_3-E^2{}_2)(E^4{}_4-E^1{}_1)|\psi_\mu\rangle\\
	&=2\langle\psi_\mu|(H_0)^2-(H_1)^2-(H_2)^2-H_0+H_1+H_2|\psi_\mu\rangle\\
	&=2\langle\psi_\mu|(d^2-j_1{}^2-j_2{}^2-d-j_1-j_2)|\psi_\mu\rangle\\
	&=2\langle\psi_\mu|(d-j_1-j_2-1)(d+j_1+j_2)|\psi_\mu\rangle
\end{align*}
for which the conditions \eqref{eq:masslessConditionsMack} are sufficient.\qed\\


\noindent\textbf{Oscillator Realisations}\\

We review here the construction of oscillator models for $L=u(2,2)\cong so(4,2)$. These can viewed as a special case of a general approach that applies to Lie algebras more broadly, see for example \cite{Todorov1966} and for Lie superalgebras \cite{Bars&Gunaydin1983}. 

We begin with the introduction of the following four-component operator-valued objects	 
\begin{align}\label{eq:oscillator_objects}
	\varphi_\pm=\left(\begin{array}{c} \pm a_1 \\ \pm a_2 \\ b^\dagger_1 \\ b^\dagger_2\end{array}\right)
	\qquad \tilde{\varphi}_\pm=\varphi_\pm^\dagger\eta_\pm=\left(a^\dagger_1,a^\dagger_2,\mp b_1,\mp b_2\right)
\end{align}
where $\eta_\pm=\pm\gamma_0=\pm diag(1,1,-1,-1)$ and the components are the bosonic creation and annihilation operators satisfying the usual relations
\[[\varphi_i,\tilde{\varphi}^j]=\delta^i{}_j\qquad\qquad [\varphi_i,\varphi_j]=[\tilde{\varphi}^i,\tilde{\varphi}^j]=0.\]  
Now for $X\in L$ and a (four-dimensional) representation $\pi:X\mapsto\pi(X)$ there corresponds two inequivalent oscillator realisations, termed $\mathcal{L}_\pm$,
\[X_+=\tilde{\varphi}_+\,\pi(X)\,	\varphi_+\qquad\mbox{and}\qquad X_-=\tilde{\varphi}_-\,\pi(X)\,	\varphi_-,\] 
where the hermiticity conditions \eqref{eq:hermiticityConditions} are satisfied as a result of \eqref{eq:oscillator_objects}. Take for example the Gel'fand generators $E^i{}_j$, $i,j=1,2,3,4$ as a basis for $su(2,2)$ satisfying \eqref{eq:comm_relns_su22_a}. In the fundamental representation each $E^i{}_j$ may be mapped to the elementary matrix $e^i{}_j$. The resulting $\mathcal{L}_+$ realisation takes the form
\[E^i{}_j= \tilde{\varphi}^l \big(e^i{}_j\big)^k{}_l \varphi_k=\tilde{\varphi}^i \varphi_j.\]
In the case of $so(4,2)$ take as a basis $J_{AB}=-J_{AB}$, $A,B = 0,1,2,3,5,6$ satisfying \eqref{eq:so42_commRels}. Using the defining representation $j_{AB}$ expressed in terms of the gamma matrices, see \eqref{eq:gammaRepSO42}, the corresponding oscillator realisations are
\[J_{AB}=\tilde{\varphi}_+\,j_{AB}\,	\varphi_+ \qquad\mbox{and}\qquad J_{AB}=\tilde{\varphi}_-\,j_{AB}\,	\varphi_-, \]
which appear in this or equivalent forms in the literature, see for example \cite{Mack&Todorov1969} and references therein. In particular the explicit evaluation of the $\mathcal{L}_-$ case appears in \cite{Wybourne} and \cite{Barut&Bohm1970}, viz
\begin{equation}\label{eq:basis_definitions}
\begin{array}{rlll}
	J_{ij}&=\frac12(a^\dagger\sigma_ka+b^\dagger\sigma_kb)\qquad\qquad\qquad&J_{56}&=\frac12(a^\dagger Cb^\dagger+aCb)\\
	J_{i5}&=-\frac12(a^\dagger\sigma_ia-b^\dagger\sigma_ib) &J_{50}&=-\frac12i(a^\dagger Cb^\dagger-aCb)\\
	J_{i0}&=-\frac12(a^\dagger\sigma_iCb^\dagger-aC\sigma_ib) &J_{06}&=\frac12(a^\dagger a + b^\dagger b+2)\\
	J_{i6}&=-\frac12i(a^\dagger\sigma_iCb^\dagger+aC\sigma_ib),\\
\end{array}
\end{equation}
where $C$ is the antisymmetric matrix\[C=\left(\begin{array}{cc}0&1\\-1&0\end{array}\right).\]
Finally Mack and Todorov \cite{Mack&Todorov1969}, see also \cite{Kursunoglu1967}, have shown that all oscillator representations of $su(2,2)$ are massless representations belonging the most degenerate series of unitary representations. See references above for a discussion of the decomposition of the Fock space into its irreducible parts and the irreducibility of oscillator representations under restriction to the Poincar\'{e} subgroup. 
 
\end{appendix}	


\end{document}